\documentclass[graybox,square]{svmult}
\usepackage{natbib}
\usepackage{slashed} 
\usepackage{mathptmx}       % selects Times Roman as basic font
\usepackage{helvet}         % selects Helvetica as sans-serif font
\usepackage{courier}        % selects Courier as typewriter font
\usepackage{type1cm}        % activate if the above 3 fonts are
\usepackage{makeidx}         % allows index generation
\usepackage{graphicx}        % standard LaTeX graphics tool
\usepackage{multicol}        % used for the two-column index
\usepackage[bottom]{footmisc}% places footnotes at page bottom

\def\gta{\mathrel{\raise.3ex\hbox{$>$}\mkern-14mu
             \lower0.6ex\hbox{$\sim$}}}
\def\lta{\mathrel{\raise.3ex\hbox{$<$}\mkern-14mu
             \lower0.6ex\hbox{$\sim$}}}

\makeindex             % used for the subject index
\begin{document}

\title*{Astrophysical Constraints on Dense Matter in Neutron Stars}
\author{M. Coleman Miller}
\institute{M. Coleman Miller \at Department of Astronomy and Joint Space-Science Institute, University of Maryland, College Park, MD 20742-2421, USA.\\ 
\email{miller@astro.umd.edu}
}

\maketitle

\abstract{Ever since the discovery of neutron stars it has been
realized that they serve as probes of a physical regime that cannot
be accessed in laboratories: strongly degenerate matter at several
times nuclear saturation density.  Existing nuclear theories
diverge widely in their predictions about such matter.  It could be
that the matter is primarily nucleons, but it is also possible that
exotic species such as hyperons, free quarks, condensates, or strange
matter may dominate this regime.  Astronomical observations of cold high-density
matter are necessarily indirect, which means that we must rely on measurements of quantities such as the masses and radii of neutron stars and
their surface effective temperatures as a function of age.  Here we
review the current status of constraints from various methods and
the prospects for future improvements.}

\section{Introduction}
\label{sec:intro}

The nature of the matter in the cores of neutron stars is
of great interest to nuclear physicists and astrophysicists
alike, but its properties are difficult to establish in
terrestrial laboratories.  This is because neutron star cores reach
a few times the density of matter in terrestrial nuclei and 
yet they are strongly degenerate and they have far
more neutrons than protons.  It thus occupies a different phase
than is accessible in laboratories.  Within current theoretical
uncertainties there are many possibilities for the state of
this matter: it could be primarily nucleonic, or dominated by
deconfined quark matter, or mainly hyperons, or even mostly in a
condensate.  

Only astrophysical observations of neutron stars can constrain the
properties of the cold supranuclear matter in their cores.  Because we
cannot sample the matter directly, we need to infer its state by
measurements of neutron star masses, radii, and cooling rates.  For
the last two of these, the method of measurement is highly indirect and thus
subject to systematic errors.  Note, to be precise, that throughout
this review we mean by mass the gravitational mass (which would be
measured by using Kepler's laws for a satellite in a distant orbit
around the star) rather than the baryonic mass (which is the sum of
the rest masses of the individual particles in the star); for a
neutron star, the gravitational mass is typically less than the baryonic mass by $\sim
20$\%.  We also mean by radius the circumferential radius,
i.e., the circumference at the equator divided by $2\pi$, rather
than other measures such as the proper distance between the stellar
center and a point on the surface.  Again, for objects as compact as
neutron stars, the difference can amount to tens of percent.

In this review we discuss current attempts to measure the relevant
stellar properties. We also discuss future prospects for
constraints including those that will come from analysis of
gravitational waves. For each of the constraint methods, we discuss the current uncertainties and
assess the prospects for lowering systematic
errors in the future. In \S~\ref{sec:nuclear} we set the
stage by discussing current expectations from nuclear theory and
laboratory measurements.  In \S~\ref{sec:mass} we examine mass
measurements in binaries.  In \S~\ref{sec:radius} we
discuss current attempts to measure the radii of neutron stars and show that most of them suffer severely from systematic errors.  
In \S~\ref{sec:cooling} we explore what can be learned from
cooling of neutron stars, and the difficulties in getting clean
measurements of temperatures.  In \S~\ref{sec:future} we
investigate the highly promising
constraints that could be obtained from the detection of
gravitational waves from neutron star--neutron star or neutron
star--black hole systems.  We summarize our conclusions in
\S~\ref{sec:summary}. For other recent reviews of equation of
state constraints from neutron star observations, see
\cite{2006PhRvC..74c5802K,2006ARNPS..56..327P,2007PrPNP..58..168S,
2007arXiv0705.2708W,2010AdSpR..45..949B,
2010ApJ...722...33S,2012ARNPS..62..485L,2013RPPh...76a6901O,
2013JPhCS.432a2001H}.

\section{Expectations from Nuclear Theory}
\label{sec:nuclear}

Any observations of neutron stars bearing on the properties of
high-density matter must be put into the context of existing
nuclear theory.  This theory, which relies primarily on laboratory
measurements of matter at nuclear density that has approximately
equal numbers of protons and neutrons, must be extrapolated
significantly to the asymmetric matter at
far higher density in the cores
of neutron stars.  We also note that the inferred macroscopic
properties of neutron stars depend on the nature of strong gravity
as well as on the properties of dense matter 
(e.g., \cite{2008PhRvD..77f4006P}), but for this
review we will assume the correctness of general relativity.

In this section we give a brief overview of current thinking about
dense matter.  We begin with simple arguments motivating the
zero-temperature approximation for the core matter and giving the
basics of degenerate matter.  We then address a  commonly-asked
question: given that the fundamental theory of quantum chromodynamics
(QCD) exists, why can we not simply employ computer calculations
(e.g., using lattice gauge theory) to determine the state of matter
at high densities?  Given that in fact such calculations are not
practical, we explore the freedom that exists in principle to
construct models of high-density matter; the fundamental point is
that because the densities are well above what is measurable in the
laboratory, one could always imagine, in the context of a model,
adding terms that are negligible at nuclear density or for symmetric
matter but important when the matter is a few times denser and
significantly asymmetric.  After discussing some
example classes of models, we survey current
constraints from laboratory experiments and future prospects.  We conclude with a discussion of how one would map an idealized future data set of
masses, radii, temperatures, etc. of neutron stars onto the equation of state of cold dense matter.

\subsection{The basics: dense matter and neutron stars}

Consider a set of identical fermions (e.g., electrons or neutrons)
of mass $m$ and number density $n$.  The linear space available to 
each fermion is thus $\Delta x\sim n^{-1/3}$, and the uncertainty
principle states that the uncertainty in momentum (and hence the
minimum momentum) is given by $\Delta p\Delta x\sim \hbar$ and
thus $p_{\rm min}\sim \hbar n^{1/3}$, where $\hbar=1.05457\times 
10^{-27}$~erg~s is the reduced Planck constant.  Done more precisely and
assuming isotropic matter we find that this minimum is the
Fermi momentum $p_F=(3\pi^2\hbar^3n)^{1/3}$.  The Fermi energy
adds to the rest-mass energy via $E_{\rm tot}=
(m^2c^4+p_F^2c^2)^{1/2}=mc^2+E_F$, and is $E_F\approx p_F^2/2m$ for
$p_F\ll mc$ and $E_F\approx p_Fc$ for $p_F\gg mc$, where
$c=2.99792458\times 10^{10}$~cm~s$^{-1}$ is the speed of light.  

\begin{figure}[t]
\includegraphics[scale=0.62]{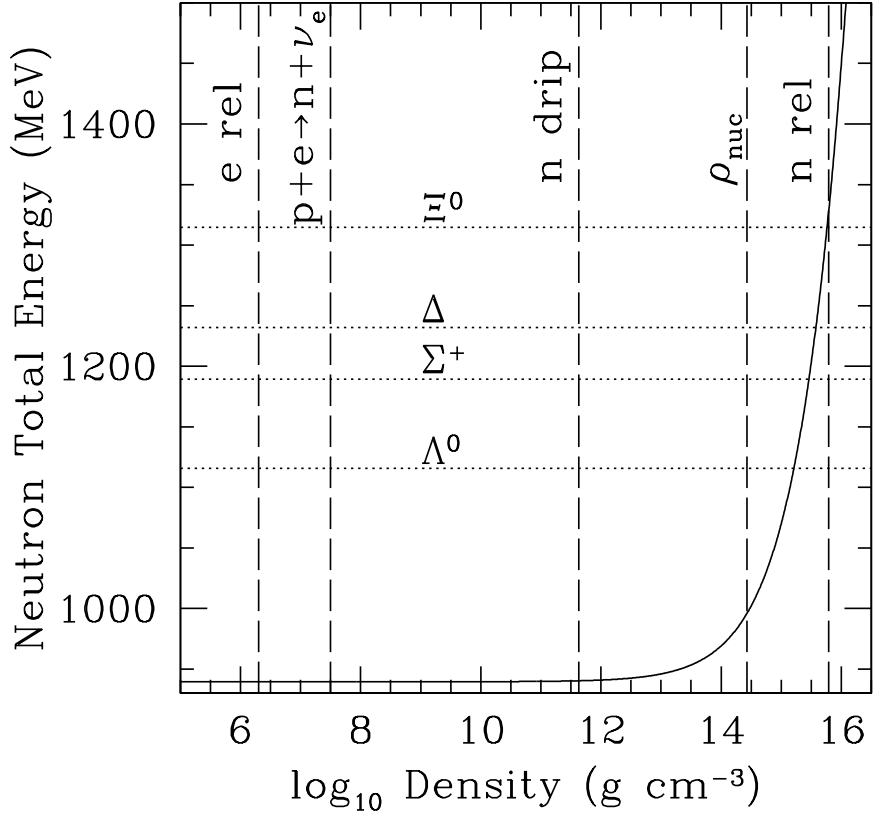}
\caption{Total energy per free neutron versus mass density
(solid line).
Above $\sim 10^{13}$~g~cm$^{-3}$ the Fermi energy starts to
contribute palpably to the total, and above $\sim 10^{15}$~g~cm$^{-3}$
the total energy can exceed the rest mass energy of particles
such as $\Lambda^0$, $\Sigma^+$, $\Delta$, and $\Xi^0$
(marked by horizontal dotted lines).  Interactions between these
particles can change the threshold density.  The central densities
of realistic neutron stars range from $\sim 5\times 10^{14}$~g~cm$^{-3}$
to $\sim{\rm few}\times 10^{15}$~g~cm$^{-3}$, so some of these
exotic particles may indeed be energetically favorable.  Also
marked are the densities at which free electrons become relativistic;
where those electrons have enough total energy to make
$p+e^-\rightarrow n+\nu_e$ possible; where free neutrons can exist
stably; nuclear saturation density; and where free neutrons have
a Fermi energy equal to their rest-mass energy.  To calculate the
neutron Fermi energy we assume that all the mass is in free neutrons;
in reality at least a few percent of the mass is in protons and
other particles, and below $\rho_{\rm nuc}$ a significant fraction
of mass is in nuclei.}
\label{fig:fermi}
\end{figure}

Matter is degenerate when $E_F>kT$, and strongly degenerate when
$E_F\gg kT$, where $k=1.38065\times
10^{-16}$~erg~K$^{-1}$ is the Boltzmann constant and $T$ is
the temperature.  As a result,
electrons (with their low masses $m_e=9.109382\times 
10^{-28}~{\rm g}=0.510999~{\rm MeV}/c^2$) become degenerate at much
lower densities than do neutrons or protons.  Matter dominated by nuclei heavier than hydrogen has 
$\sim 2$ baryons per electron, and hence electrons become relativistically
degenerate ($p_F=m_ec$) at a density of 
$\approx 2\times 10^6$~g~cm$^{-3}$.  In addition, above
$\approx 2.5\times 10^7$~g~cm$^{-3}$ the total energy of electrons
becomes larger than $m_nc^2-m_pc^2=1.294~{\rm MeV}$, where 
$m_n=1.674927\times 10^{-24}~{\rm g}=939.566~{\rm MeV}/c^2$ and 
$m_p=1.672622\times 10^{-24}~{\rm g}=938.272~{\rm MeV}/c^2$ are
respectively the rest masses of neutrons and protons.  As a result,
at these and greater densities electrons and protons can undergo inverse beta decay $e^-+p\rightarrow n+\nu_e$.
At higher densities the ratio of neutrons to protons in nuclei increases,
and then at the ``neutron drip" density $\rho_{nd}\approx
4\times 10^{11}$~g~cm$^{-3}$ neutrons are stable outside nuclei.

At infinite density, equilibrium matter consisting of just
neutrons, protons, and electrons would have eight times as many
neutrons as protons (and electrons, because charge balance has
to be maintained).  To see this, note that at infinite density
all species are ultrarelativistic and their chemical potentials
are thus dominated by their Fermi energies.  Charge balance
means that $n_p=n_e$, so equilibrium implies
\begin{equation}
\begin{array}{rl}
E_{F,n}&=E_{F,p}+E_{F,e^-}\\
&=2E_{F,p}\\
n_n^{1/3}&=2n_p^{1/3}\\
n_n&=8n_p\; .\\
\end{array}
\end{equation}

For a neutron star with a canonical
mass $M=1.4~M_\odot$ (where $M_\odot=1.989\times 10^{33}$~g is
the mass of the Sun) and radius $R=10$~km that for simplicity we
will treat as made entirely of free neutrons, the average number density
is $n=(M/m_n)/(4\pi R^3/3)=4\times 10^{38}$~cm$^{-3}$.  This
implies $p_F=2.4\times 10^{-14}$~g~cm~s$^{-1}$ and thus
$E_F\approx 2\times 10^{-4}$~erg$=100$~MeV.  This corresponds
to a temperature of $T_F=E_F/k\approx 10^{12}$~K, which is much hotter
than the expected interior temperatures $T<10^{10}$~K typical
of neutron stars more than a few years old \cite{2006NuPhA.777..497P}.  
Neutron stars are strongly degenerate.

Note, however, that the high Fermi energy of neutrons suggests
the possibility of additional particles at high densities.
For example, the lambda particle has a rest mass of
$m_\Lambda=1115.6$~MeV/c$^2$, so if the neutron Fermi energy exceeds
176~MeV then the lambda is in principle stable.  Several other
particles are within 300~MeV/c$^2$ of the neutron.  In addition,
because the density at the center of a neutron star is a few
times nuclear saturation density 
$\rho_s\approx 2.6\times 10^{14}$~g~cm$^{-3}$, quarks may
become deconfined or matter might transition to a state
that is lower-energy than nucleonic matter even at zero pressure
(strange matter; see, e.g., \cite{1984PhRvD..30.2379F}).
Various density thresholds are summarized in Figure~\ref{fig:fermi}.

Stars supported by nonrelativistic degeneracy pressure
(a reasonable approximation for neutron stars, because
$E_F<m_nc^2$) have radii that decrease with increasing mass
in contrast to most other objects.  To see this, consider
a star with a mass $M$ and radius $R$ supported by 
nonrelativistic fermions of mass $m$.  The Fermi energy per 
particle is $E_F\sim p_F^2/2m\sim \hbar^2n^{2/3}\sim
(M/R^3)^{2/3}\sim M^{2/3}/R^2$.  The gravitational energy
per particle is $E_G\sim -GMm/R$, 
where $G=6.67\times 10^{-8}$~g$^{-1}$~cm$^3$~s$^{-2}$ is
Newton's gravitational constant, so the total energy per
particle is $E_{\rm tot}=C_1M^{2/3}/R^2-C_2M/R$ where
$C_1$ and $C_2$ are constants.  Minimizing with respect to
$R$ gives $R\sim M^{-1/3}$.  Effects associated with
interactions can change
this slightly, but in practice most equations of state produce a radius that either decreases with increasing mass
or is nearly constant over a broad range in mass.  This led
\cite{2001ApJ...550..426L} to note that even for a star of
unknown mass a measurement of the radius to within $\sim 10$\%
would provide meaningful constraints on the equation of state.

\subsection{Models of matter at high densities}

There are several classes of matter beyond nuclear density:
ones in which neutrons and protons are the only baryons, ones
in which other baryons enter (especially those with strange
quarks), ones involving deconfined quark matter, and so on.
Within each class there are a number of adjustable parameters,
some of which are constrained by laboratory measurements at
nuclear density or below but many of which can be changed to
accommodate observations of neutron stars.

When confronted by this complexity a common question is: why
is there uncertainty about dense matter?  The fundamental theory,
QCD, is well-established.  Asymptotic freedom is not reached at neutron star densities, so the coupling constant
is large enough that expansions similar to those in quantum
electrodynamics are not straightforward, but in principle one could
imagine Monte Carlo calculations that establish the ground
state of degenerate high-density matter.

This approach is unfortunately not currently practical, 
due to the lack of a viable algorithm for high baryon densities.
This is because of the so-called ``fermion sign problem", which has been
known for many years.  We start by considering a representative but small volume
of matter at some density and chemical potential $\mu$ that can exchange
energy and particles with its surroundings but has a fixed
volume \cite{2010IJMPA..25...53H}.  The thermodynamic state of the matter is therefore described by a grand canonical ensemble using a partition function
\begin{equation}
Z=Z(T,\mu)={\rm Tr}\left\{\exp[-({\cal H}-\mu N)/kT]\right\}
\end{equation}
where ${\cal H}$ is the Hamiltonian and $N$ is the particle
number operator.  It is common to use $\beta\equiv 1/(kT)$.
The trace is evaluated over Fock space, which
makes this formulation inconvenient.  One can instead rewrite
$Z$ as
\begin{equation}
Z=\int {\cal D}A\, {\rm det} M(A)e^{-S_G(A)}\; ,
\end{equation}
that is, as a Euclidean functional integral over classical
field configurations.  Here $A$ represents the degrees of
freedom (quarks, gluons, $\ldots$), $S_G(A;\beta)=
\int_0^\beta dx^4\int d^3x {\cal L}_G^E(A)$ is the
thermal Euclidean gauge action, and the quark propagator matrix is
$M(A)=\slashed D(A)-m-\mu\gamma_4$ where
$\slashed D=\gamma_\mu(\partial_\mu-igA^a_\mu t^a)$,
$t^a$ are the Hermitian group generators, $g$ is the strong
coupling constant, and $\gamma$ are
the Euclidean gamma matrices.

For a vanishing chemical potential $\mu=0$, ${\rm det}M(A)$
is positive definite, meaning that all contributions add
in the same direction and $Z$ is comparatively straightforward
to compute.  If instead $\mu$ is real and nonzero, then
${\rm det}M(A)$ is complex in general.  Thus although $Z$
is still real and strictly positive, the integrands have
various phases and the integral takes the form of a cancellation
between large quantities.  The bad news is that the general
fermion sign problem is NP-hard \cite{2005PhRvL..94q0201T},
but work is proceeding on better approximation methods.
If a first-principles evaluation of
$Z$ yields, without ambiguity, the equilibrium state
of dense matter at low temperatures, then measurements
of the properties of neutron stars would serve as important
tests of QCD itself and would thus be probes of very fundamental
physics indeed.  Until that point, however, it is necessary to
use phenomenological models.

It is very difficult to rule out an entire class of models (e.g.,
those with only nucleonic degrees of freedom or those with
significant contributions from hyperons).  This is because neutron
star core densities and the asymmetry in the number densities of neutrons and protons
are significantly greater than those that can
be probed in laboratories.  As a result, one could always imagine
adding contributions that involve high powers of the density or
asymmetry.  These contributions
would have a negligible impact on laboratory matter but would have
important effects in the cores of neutron stars.  One can make some general statements: for instance, if
non-nucleonic components become important above some density the
equilibrium radius at a given mass and the maximum mass both tend
to be smaller than if only nucleonic degrees of freedom contribute
(because the presence of a new energetically favorable composition
softens the equation of state).  Unfortunately, it is  difficult to establish a
particular mass or radius that would eliminate such exotic
models.  For example, hyperonic models of neutron stars have been
constructed with maximum masses $>2.0~M_\odot$
\cite{2006PhRvD..73b4021L}.
Nonetheless, although neutron star observations cannot
entirely rule out model classes in principle, their role is
important because they probe a different realm of matter
than what is accessible in laboratories.  Ockham's razor will then be used to
judge between different model classes: if one class fits all
data using a small number of parameters that have
reasonable values and other classes require great complexity or
unreasonable values, the first class would be preferred.

\begin{figure}[t]
\includegraphics[scale=0.62]{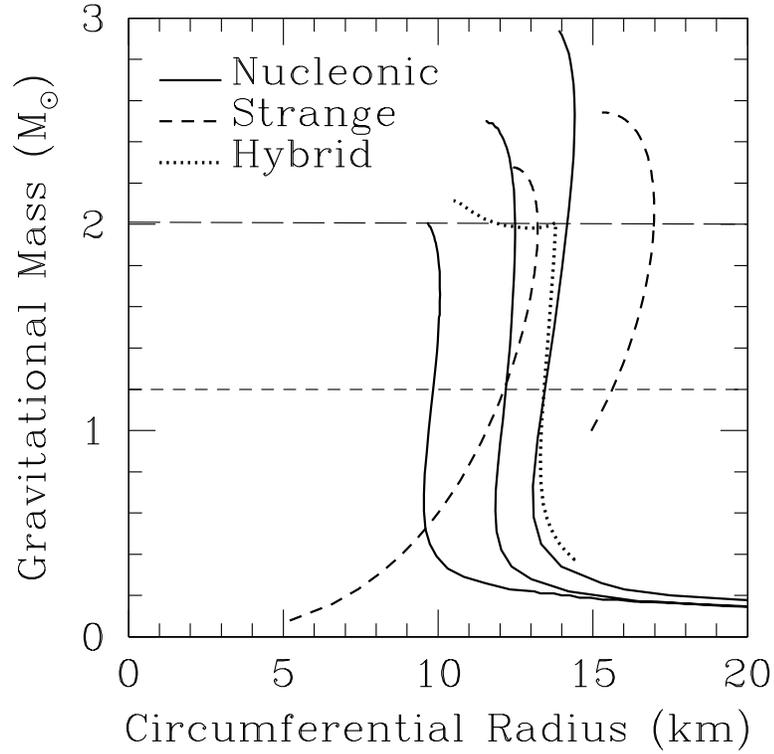}
\caption{Mass versus radius for nonrotating stars constructed
using several different
high-density equations of state.  Rotation changes the radius to
second order in the spin rate, but the corrections are minor for known neutron stars.
The solid curves include only nucleonic degrees of freedom (these are the mass-radius relations for the soft, medium, and hard equations of state from \cite{2013ApJ...773...11H}), the short dashed lines assume bare strange matter \cite{2010PhRvD..81j5021K}, and the dotted curve uses a hybrid quark equation of state with a phase transition \cite{2013arXiv1310.3803B}.   
The horizontal dashed line at $1.2~M_\odot$
represents approximately the minimum gravitational mass for a
neutron star in current formation scenarios, whereas the horizontal dashed line at $2.01~M_\odot$ shows the highest precisely measured gravitational mass for a neutron star.
}
\label{fig:mvr}
\end{figure}

One basic category of models, relativistic mean field theories, is quite
phenomenological in nature.  In these models the degrees
of freedom are nucleons and mesons (which couple minimally
to the nucleons but the coupling could have some density dependence).
The coupling strengths can be adjusted to laboratory data
and/or neutron star observations.  In a more microscopically
oriented approach, one starts instead from some given nucleon-nucleon
interaction (which can be extended to more than two nucleons)
that is fitted to data including the binding energy of light
nuclei and scattering data (for a recent effort in the
context of chiral effective field theory, see
\cite{2010arXiv1007.1746H}).  In both approaches there is
considerable freedom about the types of particles considered,
e.g., the particles could be nucleons or the particles could include hyperons or
deconfined quarks.  We plot some mass-radius relations from representative
equations of state in Figure~\ref{fig:mvr}.  It is clear that models can be constrained tightly if more massive neutron stars are discovered, or if neutron star radii can be measured with accuracy and precision (especially for stars of known mass).

Constraints on the equation of state of cold dense matter can be obtained from astronomical observations or laboratory experiments.  Some of the more useful experimental data come from relativistic heavy-ion collisions, which
can reach 2 to 4.5 times nuclear saturation density \cite{2002Sci...298.1592D}
but which have relativistic temperatures and are therefore
not degenerate.  Laboratory data also include the binding energies of
light nuclei and recent measurements of the neutron skin
thickness of heavy nuclei such as $^{208}$Pb ($0.33^{+0.16}_{-0.18}$~fm
according to the PREX team \cite{2012PhRvL.108k2502A}; see
\cite{2013arXiv1305.7101P} for some of the implications of the
expected more precise future measurements), which provide a rare glimpse of
neutron-rich matter because the neutron wavefunctions extend slightly beyond the proton wavefunctions.  These experiments thus measure the
microphysics semi-directly, whereas all astrophysical observations
place indirect constraints.  In order to make explicit contact between
microphysics and astrophysics we now discuss briefly how
to construct models of neutron stars given a high-density equation
of state.

\subsection{Construction of neutron star models from microphysics}

We argued earlier that the Fermi energy in the cores of neutron
stars is much greater than the thermal energy.  If we also assume
that the matter is in its ground state at a given density, this
implies that the pressure is only a function of density:
$P=P(\rho)$ (this is called a {\it barotropic} equation of state).
If we have such an equation of state 
we can compute the structure of a nonrotating and hence
spherically symmetric star using the
Tolman-Oppenheimer-Volkoff (TOV) equation \cite{1939PhRv...55..374O}:
\begin{equation}
{dP(r)\over{dr}}=-{G\over r^2}\left[\rho(r)+
{P(r)\over c^2}\right]\left[M(r)+4\pi r^3{P(r)\over c^2}\right]
\left[1-{2GM(r)\over{c^2r}}\right]^{-1}
\end{equation}
where $M(r)=\int_0^r 4\pi\rho(r)r^2dr$ is the gravitational
mass integrated from the center to a circumferential radius $r$.  Note that for
$P/c^2\ll\rho$ and $GM/c^2\ll r$ this reduces to the Newtonian equation of
hydrostatic equilibrium $dP/dr=-GM\rho/r^2$.  Thus one can construct
a model star by choosing a central density or pressure and integrating
to the surface, which is defined by $P=\rho=0$.  This gives the radius
and gravitational mass of the star.  Similar constructions are possible for
stars that rotate uniformly or differentially in some specified
manner (see \cite{1994ApJ...424..823C,2003LRR.....6....3S}), but 
it is conceptually clearer to focus on the nonrotating case.

One can therefore determine the neutron star mass-radius relation from a given
equation of state.  If we suppose that in the future we will have precise
measurements of the radii and gravitational masses of a large number of neutron
stars, say from the minimum possible to the maximum possible mass, then
comparison of the observed $M-R$ curve with predicted curves will strongly
constrain the parameters of a given class of models. But is it possible to go in
the other direction, that is, could one take observed $(M,R)$ pairs and infer
$P(\rho)$ directly while remaining agnostic about the microphysics that produces the equation of state?

This is not trivial.  One might imagine that the construction of
$P(\rho)$ would proceed as follows.  First, we assume that we know the equation of state up to nuclear saturation density $\rho_s$.  Even in this first step we therefore make an extrapolation from the nearly symmetric nuclear matter in nuclei to the highly asymmetric matter in neutron stars.  We use this equation of state to compute the mass $M_s$ and radius $R_s$ of a star with a central density of $\rho_s$.  We then observe a star with a slightly larger 
mass than $M_s$.  The microscopic unknowns would be the central
density (slightly larger than nuclear saturation) and the pressure
at that density, which our two measurements (of $M$ and $R$) are
sufficient to constrain.  We then bootstrap $P(\rho)$ by measuring
the mass and radius of successively more massive neutron stars.

The difficulty with this procedure is evident from Figure~11 of
\cite{1998PhRvC..58.1804A}, which  shows that a star whose central density is
exactly nuclear saturation density has a total mass of only $\sim 0.1~M_\odot$. 
To get to the $M\approx 1.2~M_\odot$ minimum for neutron star masses
\cite{1984ApJ...277..791N,2002RvMP...74.1015W}  requires densities that are more
than twice nuclear saturation.  We will thus be required to extrapolate well
beyond known matter in density and nuclear asymmetry to fit neutron star data. 
This is not fatal, because it means that we are simply comparing model predictions with
data, but it does mean that $P(\rho)$ cannot be inferred blindly without models.

If additional assumptions are made, for example that between
fiducial densities the equation of state is a polytrope
$P\propto\rho^\gamma$, then \cite{2009PhRvD..80j3003O} have shown
that precise mass and radius measurements of as few as three
neutron stars could suffice to give an empirically determined
$P(\rho)$.  The inferred $P(\rho)$ would then be compared with the predictions
from microphysical equations of state.  As discussed by 
\cite{2006NuPhA.777..479L}, there are additional relatively
model-independent constraints on the equation of state that one
can infer from observations.  For example, the typical radius of a neutron star 
scales as the quarter power of the pressure near nuclear saturation density,
and the maximum density that can be reached in a neutron star
is $\rho_{\rm max}\approx 1.5\times 10^{16}~{\rm g~cm}^{-3}
(M_\odot/M_{\rm max})^2$ if the maximum mass is $M_{\rm max}$.

Our final note about the theoretical predictions is that
there are some phenomena that have little effect on the
mass-radius relation but are important for other
observables.  For example, the existence of a proton 
superconducting gap can modify core cooling dramatically
\cite{2006NuPhA.777..497P}, but the predicted gap energies of
$\sim 1$~MeV \cite{2006NuPhA.777..497P} are so small compared
to the Fermi energy that the overall structure of neutron
stars will be affected minimally.  Thus if
neutron star temperatures and ages, particularly those of isolated
neutron stars, can be inferred reliably, then they will provide a
beautiful complement to the mass and radius measurements that
are emphasized more in this review.

\section{Constraints on Mass from Binary Observations}
\label{sec:mass}

Mass measurements of neutron stars in binaries provide the most certain of all
constraints on the properties of cold high-density matter,
particularly when the companion to the neutron star is also a neutron star and
thus the system approaches the ideal of two point masses. 
In this section we discuss such measurements,
beginning with what can be learned from purely Newtonian
observations and moving on to the greater precision and breaking
of degeneracies that are enabled by measurements of post-Keplerian
parameters from systems involving pulsars.

\subsection{Newtonian observations of binaries}

The classical approach to mass measurements in binaries assumes
that one sees periodic variation in the energy of spectral lines
from one of the stars in the binary, which we will call star 1. 
The period of variation is the orbital period $P_{\rm orb}$, the
shape of the variation gives the eccentricity $e$ of the orbit,
and the magnitude $K_1$ (which has dimensions of speed)
of the variation indicates the
line-of-sight component of the orbital speed of star 1.
Using Kepler's laws these observed quantities can be combined
to form the ``mass function", which for a circular orbit is
\begin{equation}
f_1(M_1,M_2)={K_1^3P_{\rm orb}\over{2\pi G}}
={M_2^3\sin^3 i\over{(M_1+M_2)^2}}\; .
\end{equation}
Here $M_1$ is the mass of the star being observed, $M_2$ is the mass of the other
star, and $i$ is the inclination of the binary orbit to our line of sight ($i=0$
means a face-on orbit,  $i=\pi/2$ means an edge-on orbit).  From this expression,
$f_1$ is the minimum possible mass for $M_2$; if $M_1>0$ or $i<\pi/2$ then
$M_2>f_1$.  If periodically shifting spectral lines are also observed from the
second star (and thus the binary is a so-called double-line spectroscopic binary), then the mass ratio is known and only $i$ is uncertain.  

The inclination can be constrained
for eclipsing systems.  Particular precision is possible in some extrasolar
planet observations because of the Rossiter-McLaughlin effect (in which
the apparent color of the star varies in a way dependent on inclination
as a planet transits across its disk; see \cite{1924ApJ....60...15R,
1937POMic...6..107M}).  This effect also produces a velocity offset,
which might have been seen in 2S 0921--63 \cite{2005MNRAS.356..621J}.
The inclination can
also be constrained in systems for which the companion to a compact object
just fills its Roche lobe.  This is because as the companion orbits it
presents different aspects to us, and the amplitude of variation depends on the inclination; for example, a star in a
face-on orbit looks the same to us at all phases, whereas star in an edge-on
orbit varies maximally in its aspect \cite{1978pans.proc...43A,
1990ApJ...350..386M}.  In practice this analysis is
limited to systems that have low-mass companions (because Roche lobe overflow
from a high-mass companion to a lower-mass compact object is usually unstable;
see \cite{2002apa..book.....F}) and that have transient accretion phases and 
hence have long intervals in
which there is effectively no accretion disk (because an active accretion
disk easily outshines a low-mass star and thus the binary periodicity is
very difficult to observe).  Neutron star X-ray binaries might be
less likely to be transient than black hole X-ray binaries, and their
companions tend to be much less massive and hence dimmer than the companions to black holes \cite{2002apa..book.....F}.
Thus despite the great success of this method for black hole binaries
it has found limited application for neutron star binaries.

\subsection{Post-Keplerian measurements of pulsar binaries}

The most precise measurements of the masses of neutron stars in 
binaries are made for systems in which additional parameters can
be measured.  The extreme timing precision of pulsars makes pulsar binaries especially good candidates for such measurements.  The new effects that can be measured are:

\begin{itemize}

\item Precession of the pericenter of the system, $\dot\omega$.

\item Einstein delay $\gamma$.  At pericenter, the gravitational redshift
from the system is maximized, as is the special relativistic
redshift because the orbital speed is highest at pericenter.

\item Binary orbital decay ${\dot P}_b$.  Gravitational waves are
emitted by anything that has a time-variable
quadrupole (or higher-order) mass moment.  This 
shrinks and circularizes binary orbits.

\item Shapiro delay, $r$ and $s$.  When the signal from the pulsar passes
near its companion, time dilation in the enhanced potential delays the signal compared to the arrival time of a photon in flat spacetime.  The magnitude of the delay over the orbit is characterized by the range $r$ and shape $s$ of the delay as a function of phase.  See \cite{2010arXiv1007.0933F} for a recent reparameterization of the Shapiro delay that may represent the error region better for some orbital geometries.

\end{itemize}

Using the notation of \cite{2009arXiv0907.3219F}, the dependences of these
post-Keplerian parameters on the properties of the binary are

\begin{equation}
\begin{array}{rl}
{\dot\omega}&=3\left(P_b\over{2\pi}\right)^{-5/3}(T_\odot M)^{2/3}
(1-e^2)^{-1}\\
\gamma&=e\left(P_b\over{2\pi}\right)^{1/3}T_\odot^{2/3}
m_c(m_p+2m_c)\\
{\dot P}_b&=-{192\pi\over 5}\left(P_b\over{2\pi}\right)^{-5/3}f(e)
T_\odot^{5/3}m_pm_c M^{-1/3}\\
r&=T_\odot m_2\\
s&=\sin i\\
\end{array}
\end{equation}
where $m_p$ is the pulsar mass, $m_c$ is the companion mass,
$M=m_p+m_c$ is the total mass, $T_\odot=GM_\odot/c^3=
4.925590947\mu$s, and $f(e)=(1+73e^2/24+37e^4/96)(1-e^2)^{-7/2}$.
For a given system, there are thus three Keplerian parameters that
can be measured (binary period, radial velocity, and eccentricity)
along with the five post-Keplerian parameters.  For a system
such as the double pulsar J0737--3039A/B
\cite{2003Natur.426..531B} additional quantities can
be measured.  Hence double neutron star systems
in which at least one is visible as a pulsar are superb probes
of general relativity and yield by far the most precise masses
ever obtained for any extrasolar objects. 

As discussed by \cite{2009arXiv0907.3219F}, the neutron stars with
the greatest timing precision are the millisecond pulsars. These,
however, are spun up by accretion in Roche lobe overflow systems,
and that accretion also circularizes the system to high precision. 
As a result, precession of the pericenter and the Einstein delay
cannot be measured.  The Shapiro delay, however, can be measured even for circular binaries, and because the Shapiro delay does not have classical contributions from
tides (unlike pericenter precession, for example), $r$ and $s$ can yield unbiased mass estimates.  As pointed out by Scott
Ransom, Shapiro delay measurements are likely to become more common
due to the development of very
high-precision timing for gravitational wave detection via pulsar
timing arrays.   The consequence is that currently the most constrained systems are
field NS-NS binaries, in which little mass transfer has taken place
in the system and the stars are thus close to their birth masses. In
contrast, recycled millisecond pulsars have had an opportunity to
acquire an additional several tenths of a solar mass via accretion.

Another possibility, which is described clearly by 
\cite{2009arXiv0907.3219F}, is that in high stellar density environments
such as globular clusters binary-single interactions could play a
major role.  For example, a pulsar could be recycled to millisecond
periods and then an exchange interaction could leave it in an
eccentric binary with a white dwarf or another neutron star.
Such a system would have a measurable ${\dot\omega}$ and $\gamma$,
and the neutron star could be high-mass and have excellent timing
precision.   

We do note that there are two
drawbacks to NS-WD systems in globulars compared to field NS-NS
systems.  First, although white dwarfs are small they are not point
masses to the degree that neutron stars are.  As a result, there is
a small contribution to the precession from the finite structure of
the white dwarfs.  Second, even at the high stellar densities of
globulars a comparatively large orbit is required for there to be
a significant probability of interaction.  To see this, note that
the rate of interactions for a binary of interaction cross section
$\sigma$ is $\tau^{-1}=n\sigma v$, where $n$ is
the number density of stars (typically in the core 
$n=10^{5-6}$~pc$^{-3}$) and $v\sim 10$~km~s$^{-1}$ is the velocity
dispersion.  For a system of mass $M$, the interaction cross
section for a closest approach of $a$, roughly equal to the semimajor
axis of the binary, is $\sigma\approx \pi a(2GM/v^2+a)$.  If
$M\approx 2~M_\odot$ and $n=10^5$~pc$^{-3}$, this implies 
$\tau=10^{10}$~yr when $a\approx 0.04$~AU.  This implies orbital
periods greater than a day, so dynamically formed NS-WD binaries
are systematically larger than NS-NS binaries formed in situ.
Thus longer observation times are required to achieve a
given precision.

\subsection{Dynamically estimated neutron star masses and future prospects}

For a recent compilation of dynamically estimated neutron star masses
and uncertainties, see \cite{2013arXiv1309.6635K}.  From the standpoint
of constraints on dense matter, the most important development over the
last few years has been the discovery of neutron stars with masses
$M\sim 2~M_\odot$, and possibly more.  The first such established mass
was for PSR~J1614--2230.  Demorest et al. \cite{2010Natur.467.1081D}
determined that its mass is $M=1.97\pm 0.04~M_\odot$, which they obtained
via a precise measurement of the Shapiro delay.  This measurement was
aided by the nearly edge-on orientation of the system (inclination angle
$89.17^\circ$), which increases the maximum magnitude of the delay and
produces a cuspy timing residual that is easily distinguished from any
effects of an eccentric orbit.  

The second large mass that has been robustly established belongs to
PSR~J0348+0432.  Antoniadis et al. \cite{2013Sci...340..448A} observed
gravitationally redshifted optical lines from the companion white dwarf.
The observed Doppler modulation of the energy of these lines yields a
mass ratio when combined with the modulation of the observed spin
frequency of the pulsar.  In addition, interpretation of the Balmer
lines from the white dwarf in the context of white dwarf models gives
a precise mass for the white dwarf, and indicates that the neutron star
has a mass of $M=2.01\pm 0.04~M_\odot$.  

In addition to these well-established high masses, there are hints that
some black widow pulsars (those that are currently evaporating their
companions) might have even higher masses.  This was first reported
for the original black widow pulsar PSR~B1957+20 \cite{2011ApJ...728...95V}.
For this star the best-fit mass is $M=2.40\pm 0.12~M_\odot$, but at the highest
allowed inclination and lowest allowed center of mass motion the mass
could be as low as $1.66~M_\odot$.  More recently, \cite{2012ApJ...760L..36R}
analyzed the gamma-ray black widow pulsar PSR~J1311--3430 and found a
mass of $M=2.7~M_\odot$ for simple heated light curves (but with 
significant residuals in the light curve), and no viable
solutions with a mass less than $2.1~M_\odot$.  They conclude that
better modeling and more observation is needed to establish a reliable
mass, but it is an intriguing possibility that black widow pulsars have
particularly large masses.

From the astrophysical standpoint, it has been proposed that neutron
star birth masses are bimodal, depending on whether the core collapse
occurs due to electron capture or iron core collapse \cite{2010ApJ...719..722S}.
There is also mounting evidence for systematically higher masses in
systems that are expected to have had substantial accretion
\cite{2011A&A...527A..83Z}.  From the standpoint of nuclear physics,
\cite{2010arXiv1012.3208L} point out that $2.0~M_\odot$ neutron stars
place interesting upper limits on the physically realizable energy
density, pressure, and chemical potential.  Higher masses
would present even stronger constraints.

To a far greater extent than with the other constraints described in
this review, we can be confident that the mere passage of time will
greatly improve the mass measurements, and indeed all of the timing
parameters.  Table~II of \cite{1992PhRvD..45.1840D} shows that as
a function of the total observation time $T$ (assuming a constant
rate of sampling), the fractional errors in the post-Keplerian
parameters scale as $\Delta{\dot\omega}\propto T^{-3/2}$,
$\Delta\gamma\propto T^{-3/2}$, $\Delta{\dot P}_b\propto
T^{-5/2}$, $\Delta r\propto T^{-1/2}$, and $\Delta s\propto
T^{-1/2}$; for the $r$ and $s$ parameters the improvements are
simply due to having more measurements, whereas the others improve
faster with time because the effects accumulate.  Particularly
good improvement is expected for
the NS-WD systems because as we describe above they have larger orbits 
and thus slower precession than NS-NS systems.  There is thus reason to hope that additional high-mass systems will be discovered.

There are also planned observatories and surveys that will dramatically
increase the number of known pulsars of all types, which will likely
include additional NS-NS and NS-WD systems.  An example of such a
planned observatory is the Square Kilometer Array,
which has been projected to increase
our known sample of pulsars by a factor of $\sim 10$.  In addition,
as \cite{2010arXiv1006.4665T} pointed out recently, future
high-precision astrometry will be able to deconvolve the parallactic,
proper, and orbital motion of the two components of a high-mass
X-ray binary.  They estimate that for parameters appropriate to the
Space Interferometry Mission this will yield a neutron star
mass accurate to 2.5\% in X~Per, to 6.5\% in Vela~X-1, and to $\sim$10\%
in V725 Tau and GX~301--2.

It is thus
probable that in the next $\sim 10$ years we will have far more,
and far better, estimates of the masses of individual neutron stars.
We do not, however, have a guarantee that any of those masses will
be close to the maximum allowed.  We thus need additional ways to
access the properties of high-density matter.  In particular, many
equations of state imply similar maximum masses but widely different
radii.  We therefore turn to constraints on the radius.

\section{Constraints on Radius, and Other Mass Constraints}
\label{sec:radius}

As discussed in \S~\ref{sec:nuclear}, accurate measurements of the
radii of neutron stars would strongly constrain the properties of
neutron star core matter.  Unfortunately, all current inferences of
neutron star radii are dominated by systematic errors, and hence no
radius estimates are reliable enough for such constraints.  However, future
measurements using the approved mission NICER
\cite{2012SPIE.8443E..13G} and the proposed missions LOFT
\cite{2012ExA....34..415F}, AXTAR \cite{2011arXiv1109.1309R}, and
ATHENA+ \cite{2013arXiv1306.2307N} hold great promise for precise
radius measurements if the effects of systematic errors can be shown to be small.  In this
section we discuss various proposed methods for measuring radii and
the diverse results obtained by applying these methods.  Some of the
methods also result in mass estimates, so we discuss those
implications along the way.

\subsection{Thermonuclear X-ray bursts}

More than thirty years ago it was proposed that the
masses and radii of neutron stars could be obtained
via measurement of thermonuclear X-ray bursts 
\cite{1979ApJ...234..609V}.  These
bursts occur when enough hydrogen or helium (or carbon
for the long-lasting ``superbursts") accumulates on
a neutron star in a binary.  Nuclear fusion
at the base of the layer becomes unstable, leading to
a burst that lasts for seconds to hours.  For a selection
of observational and theoretical papers on thermonuclear
bursts, see
\cite{1976ApJ...205L.127G,1976ApJ...206L.135B,
1976Natur.263..101W,1977Natur.270..310J,1978ApJ...220..291L,
1978ApJ...224..210T,1982ApJ...261..332F,
1987ApJ...323L..55F,1993ApJ...413..324T,1993SSRv...62..223L,
2002ApJ...566.1018S,2005ApJ...629..422C}.
In some bursts, fits of a Planck function to the spectra
reveal a temperature that initially increases, then
decreases, then increases again before finally decreasing
\cite{1976MNRAS.177P..83L,1980ApJ...240L..27H,2001ApJ...562..957S}.
These are called photospheric radius expansion (PRE) bursts.
The usual assumption is that PREs occur because the
radiative luminosity exceeds the Eddington luminosity
\begin{equation}
L_E={4\pi GMc\over{\kappa}}
\end{equation}
where $M$ is the mass of the star and $\kappa$ is the radiative
opacity.  At luminosities greater than  $L_E$, an optically 
thick wind can be driven a
potentially significant distance from the star.  This leads to an
increase in the radiating area and a consequent decrease in the
temperature \cite{1983ApJ...267..315P,1984PASJ...36..551E}.   For
Thomson scattering in fully ionized matter with a hydrogen mass
fraction $X$,  $\kappa=0.2(1+X)$~cm$^2$~g$^{-1}$ and thus
\begin{equation}
L_E=2.6\times 10^{38}~{\rm erg~s}^{-1}(1+X)^{-1}(M/M_\odot)\; .
\end{equation}

The basic method of \cite{1979ApJ...234..609V} involves several
assumptions.  These are:

\begin{enumerate}

\item The full surface radiates uniformly after the photosphere has retreated to the radius of the star.

\item The stellar luminosity is the Eddington luminosity at the point of ``touchdown", which is defined as the time after the peak inferred photospheric radius is reached when the color temperature, derived from a Planck fit to the spectrum, is maximal.  The luminosity can then be determined via measurement of the flux at Earth and the distance to the star, assuming that the flux is emitted isotropically.

\item The spectral model for the atmosphere is correct.  Thus the model must have been verified against data good enough to distinguish between models, and the atmospheric composition must be known.  It is typically assumed that the color factor $f_c\equiv T_{\rm col}/T_{\rm eff}$, which is the ratio between the fitted Planck temperature and the effective temperature, is not only known but is constant throughout the cooling phase.

\item All other sources of emission from the system are negligible.

\end{enumerate}

Using these assumptions, and using the notation of \cite{2010ApJ...722...33S},
if we have measured the distance $D$ to the star and know 
$\kappa$, we can measure the touchdown flux
\begin{equation}
F_{\rm TD,\infty}={GMc\over{\kappa D^2}}\sqrt{1-2\beta(r_{\rm ph})}
\end{equation}
where $\beta(r)\equiv GM/rc^2$, the factor before the square root is the Eddington flux diluted
by distance, and $r_{\rm ph}$ is the radius of the photosphere.
We can also use the cooling phase of the burst to define a normalized
angular surface area
\begin{equation}
A={F_\infty\over{\sigma_{\rm SB} T_{\rm col,\infty}^4}}=f_c^{-4}\left(R\over D\right)^2
(1-2\beta)^{-1}\; .
\end{equation}
Here $\sigma_{\rm SB}=5.6704\times 10^{-5}$~erg~cm$^{-2}$~s$^{-1}$~K$^{-4}$ is the Stefan-Boltzmann constant and $F_\infty$ and $T_{\rm col,\infty}$ are the flux and fitted Planck temperature that we measure in the cooling phase.  Then the combinations of observed quantities
\begin{equation}
\begin{array}{rl}
\alpha&\equiv {F_{\rm TD,\infty}\over{\sqrt{A}}}{\kappa D\over{c^3 f_c^2}}\\
\gamma&\equiv {Ac^3f_c^4\over{F_{\rm TD,\infty}}\kappa}\\
\end{array}
\end{equation}
can be related to $\beta$ and $R$ by $\alpha=\beta(1-2\beta)$ and
$\gamma=R[\beta(1-2\beta)^{3/2}]^{-1}$ and solved to yield
\begin{equation}
\begin{array}{rl}
\beta&={1\over 4}\pm {1\over 4}\sqrt{1-8\alpha}\; ,\\
R&=\alpha\gamma\sqrt{1-2\beta}\; ,\\
M&=\beta Rc^2/G\; .\\
\end{array}
\end{equation}

The problem is that for several bursters, the most probable values of the observationally inferred quantities $F_{{\rm TD},\infty}$, $A$, and $D$, combined with the model parameter $f_c$, yield $\alpha>1/8$.  This would imply that the mass and radius are complex numbers.  For example, in their analysis of 4U~1820--30, \cite{2010ApJ...719.1807G} used a Gaussian prior probability distribution for $F_{{\rm TD},\infty}$, which had a mean $F_0=5.39\times 10^{-8}$~erg~cm$^{-2}$~s$^{-1}$ and a standard deviation $\sigma_F=0.12\times 10^{-8}$~erg~cm$^{-2}$~s$^{-1}$.  They also used a Gaussian prior probability distribution for $A$, with $A_0=91.98~({\rm km}/10~{\rm kpc})^2$ and $\sigma_A=1.86~({\rm km}/10~{\rm kpc})^2$.  Their prior probability distribution for $D$ was a boxcar distribution with a midpoint $D_0=8.2$~kpc and a half-width of $\Delta D=1.4$~kpc.  Finally, they assumed a boxcar prior probability distribution for the color factor, with $f_{c0}=1.35$ and a half-width $\Delta f_c=0.05$.  If we take the midpoint of each distribution and also follow \cite{2010ApJ...719.1807G} by assuming that the opacity is dominated by Thomson scattering and thus $\kappa=0.2$~cm$^2$~g$^{-1}$ for the pure helium composition appropriate to 4U~1820--30, we find $\alpha=0.179>1/8$.
 
\cite{2010ApJ...719.1807G} note that the probability of obtaining a viable $M$ and $R$ for 4U~1820--30 from these equations drops with increasing distance, but if we reduce $D$ to the 6.8~kpc, which is the smallest value allowed in the priors of \cite{2010ApJ...719.1807G}, and keep the other input parameters fixed, we find $\alpha=0.148$.  If we also increase $f_c$ to its maximum value of $1.4$ and take the $+2\sigma$ value of $A$ and the $-2\sigma$ value of $F_{{\rm TD},\infty}$, $\alpha$ is still $1.29$.  In fact  \cite{2010ApJ...722...33S} showed that if we consider the prior probability distribution of $F_{\rm TD,\infty}$, $A$, $D$, and $f_c$ used by \cite{2010ApJ...719.1807G}, only a fraction $1.5\times 10^{-8}$ of that distribution yields real numbers for $M$ and $R$.  This demonstrates that the 4\% fractional uncertainties on the mass and radius of this neutron star obtained by \cite{2010ApJ...719.1807G} emerge from the theoretical assumptions rather than from the data.  Thus such apparent precision is actually a red flag that one or more of the model assumptions is incorrect.

The first suggestion for which assumption is in error came from \cite{2010ApJ...722...33S}, who proposed that although the entire surface still
emits uniformly throughout the cooling phase, the photospheric radius
might be larger than the radius of the star, i.e., $r_{\rm ph}>R$.  
However, analysis of the cooling phase of the superburst from 4U~1820--30
\cite{2011fxts.confE..24M} demonstrates that such a solution is
disallowed for at least the superburst emission from this star, 
because any detectable change in photospheric radius would require a flux
very close to Eddington, and such fluxes give extremely poor fits to the
data.  The work of \cite{2011fxts.confE..24M} used and verified
the fully relativistic Comptonized spectral models of
\cite{2012A&A...545A.120S}, and also showed that the {\it fraction} of the
surface that emits changes systematically throughout the superburst (the
emitting area drops by $\sim 20$\% during the $\sim$1600 seconds analyzed). 
Moreover, there is no guarantee that the whole surface was emitting at {\it any} time.
Thus the star does not emit uniformly over its entire surface during the
superburst, and hence it cannot be assumed that it has uniform emission
during shorter bursts when the data quality is insufficient to check
this assumption.  Indeed, the presence of burst oscillations (see
\cite{2012ARA&A..50..609W} for a recent review) demonstrates that there is
nonuniformity in burning during many bursts.  Additional
concerns are that the color factor is likely to evolve during the burst, and that some of the bursts are not fit well using existing spectral models \cite{2010ApJ...720L..15B,2011ApJ...742..122S,2012MNRAS.422.2351C,2012arXiv1204.3486Z,2012ApJ...747...75G,2013MNRAS.429.3266G}.

Suleimanov et al. \cite{2011ApJ...742..122S} find a radius of more
than 14~km from a long PRE burst from 4U~1724--307 based on fits of
their spectral models to the decaying phase of the burst.  As part
of their fits, they find that the Eddington flux occurs not at
touchdown, but at a 15\% lower luminosity; this, therefore, calls
into question another of the assumptions in the standard approach
to obtaining mass and radius from bursts.  Their radius value is
based on the good fit they get to the bright portion of the burst,
when according to their fits the local surface flux exceeds $\sim 50$\% of 
Eddington.  This is an intriguing method that should be considered
carefully when data are available from the next generation of X-ray
instruments.  However, a potential concern is that the spectra
does not agree with their models below $\sim 50$\% of Eddington.
This suggests that there is other emission in the system at least
at those lower luminosities, and hence that some of this emission might
be present at higher luminosities as well.  Against this is the
excellent fit without extra components that
\cite{2011fxts.confE..24M} obtained for the superburst from
4U~1820--30 using the models of \cite{2012A&A...545A.120S}.  More
and better data are the key.

If truly excellent spectra can be obtained, then as pointed out by
\cite{2005AcA....55..349M}, inference of the surface gravity $g$ and
surface redshift $z$ leads uniquely to determination of the stellar
mass and radius.  However, even the $\sim 2\times 10^7$ counts
observed using {\it RXTE} from the 4U~1820--30 superburst are insufficient to
determine both $g$ and $z$ uniquely \cite{2011fxts.confE..24M}, so this appears
to require much larger collection areas.  The combination $(1+z)/g^{2/9}$
can be measured precisely using sufficiently good continuum spectra,
and then combined with other measurements to, possibly, constrain
$M$ and $R$ \cite{2011fxts.confE..24M,2013ApJ...776...19L}, so this
is promising for the future.  The net result is that currently
inferred  masses and radii from spectral fits to thermonuclear X-ray bursts must be
treated with caution; none are currently reliable enough to factor
into equation of state constraints.

\subsection{Fits of thermal spectra to cooling neutron stars}

In principle, observations of cooling neutron stars with known 
distances allow us to measure the radii of those stars, modulo an
unknown redshift.  In practice, as with radius estimates from
bursts, systematic errors dominate and thus current radius
determinations are not reliable enough to help constrain the
properties of dense cold matter.

To understand the basic principles, suppose that the star is at
a distance $d$ and that we measure a detector bolometric flux $F_{\rm
det,bol}$ from the star that is fit by a spectrum with an effective
temperature $T_{\rm eff,\infty}$ at infinity.  Suppose that we also
assume that the entire surface radiates uniformly and
isotropically.  If the surface redshift is $z$ then the luminosity
at the surface is $L_{\rm surf}=(1+z)^2L_\infty=(1+z)^2F_{\rm
det,bol}4\pi d^2= 4\pi R^2 \sigma_{\rm SB} T_{\rm surf}^4= 4\pi R^2
(1+z)^4\sigma_{\rm SB} T_{\rm eff,\infty}^4$.  This implies
\begin{equation}
R=(1+z)^{-1}d[F_{\rm det,bol}/(\sigma_{\rm SB} T_{\rm eff,\infty}^4)]^{1/2}\; .
\end{equation}
The key questions are therefore (1)~how well can the distance be
determined, (2)~how well can $T_{\rm eff,\infty}$ be established,
given that it depends on a spectral fit, and (3)~how well is
the flux known, since only the thermal component is relevant?

There are two categories of sources that have been studied carefully
in this manner to estimate neutron star radii. The first is the
so-called quiescent low-mass X-ray binaries (qLMXBs).  These are
transiently accreting neutron stars that, in the ideal case, do not
accrete at all when they are not accreting actively.  Some qLMXBs
are in globular clusters, so for those
sources the distance to the qLMXBs can be determined by measuring the
distance to the cluster.  If a qLMXB has a phase of active accretion,
then it has a steady supply
of hydrogen or helium, which should rise to the top and dominate the
surface composition after a few seconds \cite{1980ApJ...235..534A}.  
In addition, the dipolar magnetic field strengths of LMXBs are weak,
typically $\sim 10^8-10^9$~G averaged over the surface \cite{2008AIPC..983..533B}.  Thus
magnetic effects will be minimal (note that magnetic fields can affect energy spectra significantly at photon energies comparable to or less than the electron cyclotron energy $\hbar\omega_B=\hbar eB/m_ec=11.6~{\rm keV}(B/10^{12}~{\rm G})$).  Therefore,
the spectra might be well-modeled by nonmagnetic atmospheres.  These
characteristics are all good for estimating radii.  
\cite{2013ApJ...772....7G} recently applied this method to five qLMXBs in globular clusters.  They assumed that all the stars have the same radius, that the atmospheres are nonmagnetic and composed purely of hydrogen (an assumption that they supported on the basis of the companion type in two
cases), and that the surface emission was uniform.  With these assumptions, they found $R=9.1^{+1.3}_{-1.5}$~km.

However, the atmospheric composition matters greatly.  For example,
\cite{2012MNRAS.423.1556S} find that whereas fits of hydrogen
atmospheres to a qLMXB in the globular cluster M28 yield a radius
of $R=9\pm 3$~km at 90\% significance, a helium atmosphere gives
$R=14^{+3}_{-8}$~km with an equally good fit ($\chi_\nu/{\rm dof}=
0.87/141$ for the H atmosphere, 0.88/142 for the He atmosphere).
Similarly, \cite{2013ApJ...764..145C} fit data from a qLMXB in
the globular cluster M13 and find $R=9.0^{+3.0}_{-1.5}$~km for
an H atmosphere and $R=14.6^{+3.5}_{-3.1}$~km for a He atmosphere,
again with a comparable quality of fit.  Although one might argue
\cite{2013ApJ...772....7G} that if the companion is hydrogen
rich the neutron star atmosphere will be as well, it has been
proposed that after $\sim 10^{2-4}$ years diffusive
burning of hydrogen will leave helium as the main surface
composition (\cite{2003ApJ...585..464C,2010ApJ...723..719C}; see
\cite{2013arXiv1307.6060B} for the first step in a re-evaluation
of this work in the light of a better treatment of Coulomb
separation of ions).  Given that $\sim 15$ years of RXTE observations
have not revealed any accretion-powered outbursts from any of
the qLMXBs included in the analysis of \cite{2013ApJ...772....7G}, the recurrence time could be significantly greater
than a century and thus the surface composition might be helium
rather than hydrogen (see \cite{2013arXiv1305.3242L} for a
discussion of what the larger implied radii would mean for
dense matter).

Moreover, it is not necessarily valid to assume that the entire
surface radiates uniformly.  Nonaxisymmetric emission could theoretically show up as detectable pulses, but if the magnetic pole
is close to the rotational pole (as was suggested to explain
phenomena including the lack of pulsations in most LMXBs; see
\cite{2009ApJ...705L..36L,2009ApJ...706..417L}) the pulsations
could be undetectable while leading to inferred radii that are
too low.

In addition, there are several puzzling aspects to the quiescent
emission.  For stars that have undergone several outbursts we have
a decent estimate of their overall accretion rate.  We can thus
estimate the luminosity produced in the crust by electron capture
reactions (e.g., \cite{2006isna.confE..46B}),  which should be the
minimum emergent luminosity.  We also expect that the flux we see
from the star should decline smoothly after cessation of
accretion, and thus not have short timescale increases and decreases.  In
addition, if cooling is the only process, the spectrum should be
purely thermal.  All these expectations are violated in several
stars. The quiescent thermal luminosity of H1905+00 is $L<2\times
10^{30}$~erg~s$^{-1}$, which is significantly below standard predictions
\cite{2007ApJ...665L.147J}.  Several stars vary in intensity by a
factor of a few on timescales of days to years during their decline
(for a discussion see \cite{2001MNRAS.325.1157U}).  In addition,
many of the field low-luminosity stars have significant, even
dominant, nonthermal tails  (e.g., Cen X-4;
\cite{2001ApJ...551..921R}).  

It is possible that these
discrepancies can be understood within the basic picture.  For
example, enhanced neutrino emission in the crust would not be
observed, so this could explain the underluminous stars.  Short
timescale variability might be caused by the motion of obscuring matter in
these binaries.  Nonthermal spectral tails could be caused by coronal
emission from the companion star
\cite{2000ApJ...541..849C,2000ApJ...541..908B}, magnetospheric
activity, or some small residual accretion. Thus although the simple
model is contradicted, plausible additions could rescue it.  This
nonetheless contributes an additional note of caution to radius 
inferences from these stars.

The second category of source for spectral modeling and radius
inferences is isolated neutron stars.  We will encounter these
again in \S~\ref{sec:cooling} when we discuss cooling
processes, but here we focus on what can be learned about radii. 
Exactly the same principles apply as for the qLMXBs, except that
these stars do not accrete (the accretion rate from the
interstellar medium is negligible; see \cite{1995ApJ...454..370B}
for a discussion).   The same questions apply for these stars as
they do for qLMXBs.  For example, the spectra of young isolated 
neutron stars (such as the Crab pulsar) have significant nonthermal
components probably caused by magnetospheric emission. Distances
are also often difficult to establish, and the spectra can be fit
using various spectra that give significantly different answers for
the radius.

A case in point is RX~J1856.5--3754, which is the brightest of the eight
thermally emitting isolated neutron stars discovered using ROSAT. 
Its spectrum is featureless and can be fitted using a Planck
spectrum, a heavy element spectrum, or a spectrum appropriate for
condensed matter with a thin hydrogen envelope
\cite{2010ApJ...722...33S}.  An interesting puzzle that constrains
the atmospheric model is that the best-fit Planck spectrum to the
X-ray data underpredicts the optical flux by a factor of $\sim 6$
\cite{2001A&A...379L..35B}. The distance has been estimated to be
$117\pm 12$~pc \cite{2002ApJ...576L.145W} or $161\pm 12$~pc
\cite{2007Ap&SS.308..191V}, although the data for the latter were
re-analyzed to find a distance of $123^{+11}_{-15}$~pc 
\cite{2010arXiv1008.1709W}.  It has been argued that this star
has a radius of $>14$~km, which if true would make a strong case
for hard equations of state.  However, better atmospheric modeling
suggests a radius of $11.5\pm 1.2$~km \cite{2010ApJ...722...33S} assuming
a distance of 120~pc, which makes the radius consistent with standard
nucleonic and quark matter equations of state.

\begin{figure}[t]
\includegraphics[scale=0.62]{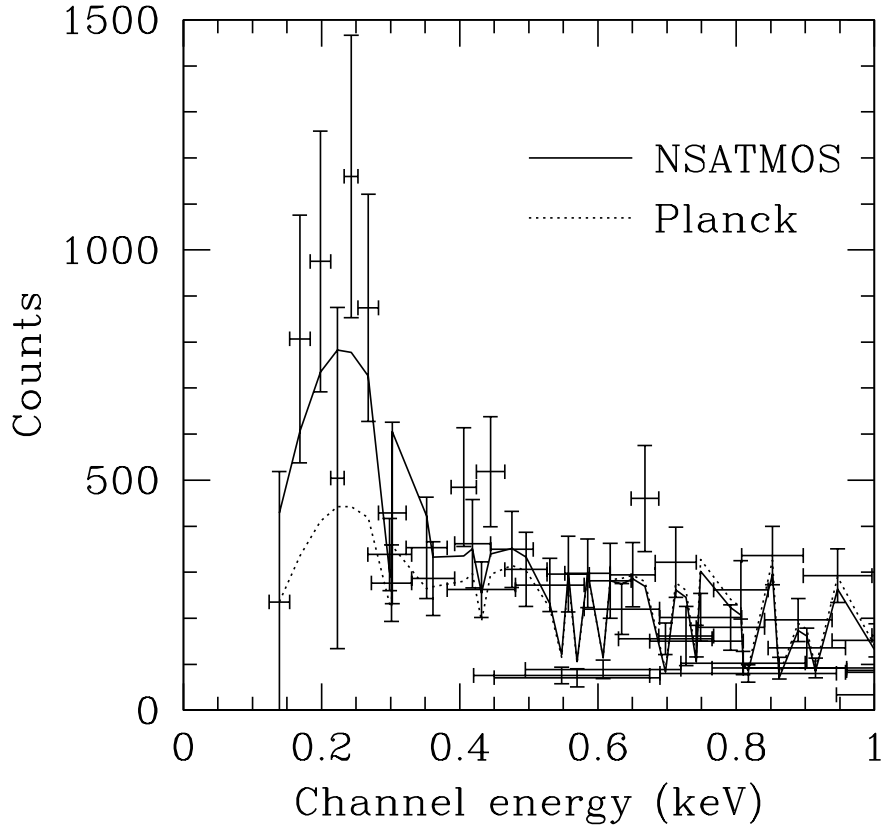}
\caption{XMM spectrum (error bars) and fits to a neutron star
atmosphere model (solid line) and a blackbody (dotted line) for a
neutron star in the globular cluster M13 (see
\cite{2007ApJ...671..727W}).  The total chi squared for the blackbody
fit is $\chi^2=88$ for 59 degrees of freedom, compared with
$\chi^2=64$ in the NSATMOS model, so for the three extra parameters the
difference is slightly greater than $4\sigma$. The fitted radius of
emission in the blackbody model is only $\sim 3$~km, which is
unphysically low unless only a small portion of the polar caps is
emitting.  The blackbody model can thus be tentatively ruled out on
physical grounds and by a goodness of fit measure.  However, a Planck
function with an efficiency much less than the 100\% efficiency of a
blackbody is viable.  Note  also that the differences
between a nonmagnetic hydrogen atmosphere and other candidates (e.g.,
pure helium, heavy atmospheres, or condensed surfaces; see
\cite{2003ApJ...585..464C,2010ApJ...722...33S}) are much less than
their differences from a blackbody, and as these give  significantly
different inferred stellar radii, caution is essential in
these inferences. Data and model kindly provided by Natalie Webb.}
\label{fig:spect}
\end{figure}

Other stars have been fit in the same manner; see for example
\cite{2007ApJ...671..727W,2009MNRAS.392..665G,2010arXiv1007.2415G}. 
Table~2 of \cite{2007ApJ...671..727W} gives fitted radii for three
neutron stars in globular clusters, and Table~4 of
\cite{2009MNRAS.392..665G} gives fits for $R_\infty\equiv
R(1-2GM/Rc^2)^{-1/2}$ for a number of qLMXBs and neutron stars in
globulars.  The uncertainties are large in many cases and thus these
measurements tend not to be good discriminants between equations of
state.  In addition, it must be kept in mind that, as with the qLMXBs, 
the stars are
so dim that their observed spectra cannot easily discriminate
between quite different models, whether these be hydrogen
atmospheres, helium atmospheres, Planck spectra, or
heavy elements (see Figure~\ref{fig:spect}). Finally, it must be
recalled that we know of nonaccreting neutron stars that pulse in
the X-rays. If the isolated neutron stars we see that do not appear
to pulse simply have their magnetic axes nearly aligned with their
rotational axes, then single-temperature fits are likely to be
misleading. Large-area instruments or long observations such as
those planned for NICER \cite{2012SPIE.8443E..13G} or LOFT
\cite{2012ExA....34..415F} would help greatly in distinguishing
between models.

\subsection{Modeling of waveforms}

Thermonuclear bursts that display brightness oscillations, and
isolated millisecond pulsars that are detectable in X-rays, have
had their periodic waveforms analyzed in attempts to 
constrain their masses and radii.  The basic principle is simple:
if one makes the standard assumption that the oscillations are
rotational modulations of the flux produced by a hot spot fixed on the star, then the
shape of the waveform encodes information about the mass and
radius.  For example, a star of a given rotation frequency that
has a large radius will have a higher linear surface rotational
speed at a given latitude than a star with a small radius.  Thus
the waveform will have greater asymmetries produced by
Doppler effects if the star is large than if it is small.
The mass to radius ratio affects the fraction of the cycle
when the spot is visible; in the Newtonian limit $R\gg GM/c^2$
exactly half the surface is visible, but for radii appropriate
to neutron stars light-bending effects make more of the surface
visible, and for a slowly rotating star with $R<3.5GM/c^2$ the entire surface can be seen.  More compact stars will therefore tend to produce lower amplitude waveforms.
Hence in principle a waveform can be analyzed to infer the
radius and mass, as well as other quantities such as the rotational
latitude of the spot center, the rotational latitude of the observer, the spot
angular radius (which turns out to be unimportant, as does the spot shape,
as long as the spot radius is significantly less than the latitude
of the spot center), and the emission pattern from the surface
as seen in the spot's local rest frame.

Work by \cite{2003ApJ...595.1066M} initially appeared to cast doubt
on this model of burst oscillations, because when they stacked data
from multiple bursts from a given star they seemed to find that
in some cases high energy photons arrive after low energy photons, i.e., the hard photons lag the soft photons.  This
is unexpected in the rotating hot spot model because as the spot
rotates into view, the Doppler effect blueshifts the spectrum and thus
high energies should lead low energies.  However, a careful re-examination
of the data for 4U~1636--536, which showed the strongest
hard-lag trend in \cite{2003ApJ...595.1066M}, showed that in fact the
oscillation phase versus photon energy is entirely consistent with
the rotating hot spot model \cite{2013MNRAS.433L..64A}.  Statistical fluctuations, and probably
the consequences of stacking burst data, appear to have led to the
opposite conclusion in \cite{2003ApJ...595.1066M}.  Thus currently
it does appear that rotating hot spots are consistent with the
data on burst oscillations.

The most common assumption in such modeling is the ``Schwarzschild
plus Doppler" approximation (e.g., \cite{1998ApJ...499L..37M,
2001ApJ...546.1098W,2002ApJ...564..353N,2005ApJ...619..483B}), in
which the star is assumed to be spherical and the spacetime
external to the star is assumed to be Schwarzschild (nonrotating),
but all the special relativistic effects associated with the rotation of the surface are treated exactly.  Full simulations that trace rays in the numerical spacetimes appropriate for rotating objects have
demonstrated that the waveforms generated using the Schwarzschild+Doppler approximation are indistinguishable from the waveforms in the full simulation when there are $\lta 10^5$ total counts and the stars have
radii $R<15$~km and spin frequencies $\nu<600$~Hz
\cite{2000ApJ...531..447B,2007ApJ...654..458C}. The current best
analyses of isolated millisecond pulsars and bursting stars yield radii that are
consistent with expectations but not very constraining;
\cite{2008ApJ...689..407B} find that J0030+0451 has a radius
$>9.4$~km  and J2124--3358 has a radius $>7.8$~km (both at 68\%
confidence, and both assuming $M=1.4~M_\odot$), and
\cite{2005ApJ...619..483B} find $Rc^2/GM>4.2$ at 90\% confidence
for the burster XTE J1814--338.  Analyses using the
``oblate Schwarzschild" approximation (in which the star is allowed
to be oblate due to rotation but the spacetime is still assumed to
be Schwarzschild) for SAX J1808.4--3658 \cite{2011ApJ...726...56M}
and XTE J1807--294 \cite{2011ApJ...742...17L} are similarly
unconstraining.  The strongest current constraints from this method
come from a recent analysis of PSR J0437--4715 assuming a hydrogen
atmosphere, for which the result is $R>11.1$~km at $3\sigma$ confidence \cite{2013ApJ...762...96B}.  

A key assumption in the analysis of \cite{2013ApJ...762...96B} is that the angular distribution of radiation from a point on the surface can be described by the pattern that emerges when the energy is deposited deep and propagates through a pure nonmagnetic hydrogen atmosphere.  This could be a correct assumption, but in addition to our previous comments that hydrogen might not be the dominant surface composition, we note that the assumption of deep deposition of energy (which for this source comes from the return current of relativistic pairs from the magnetosphere) is based on the idea that the current gives up its energy via bremsstrahlung and nothing else.  Given that plasma instabilities can shorten by orders of magnitude the column depth of energy deposition (see \cite{2012ApJ...752...22B} for a recent example in the context of how AGN jet energy is injected in the intergalactic medium), this might not be a safe assumption.

There are two reasons for the large confidence
regions that currently arise from analyses of waveforms: (1) the total number of counts is small, and (2) there
are significant degeneracies between the parameters that produce
the waveform.  The effects of both factors are expected to be addressed using the
next generation of large-area X-ray timing satellites.  As
discussed by \cite{2013ApJ...776...19L}, if a million counts are
received from the spot (comparable to the total number expected
from combining several bursts observed using LOFT) and the center of the hot
spot and the observer are both within $10^\circ$ of the rotational
equator, then $M$ and $R$ can both be obtained to 10\% precision.

As with the other methods we discuss, a key question is the role
of systematic errors: if some aspect of the real system differs
from what we assume in our model fits, how badly will our
inferred mass and radius be skewed?  There are clearly an unlimited
number of possible sources of systematic error, but an encouraging
conclusion from the work done by 
\cite{2013ApJ...776...19L} with synthetic data is that even if the assumed surface beaming pattern, spot shape, or spectrum differ significantly from the actual ones, fits
using the standard model do not {\it simultaneously} produce
(1)~a statistically good fit, (2)~apparently strong constraints
on $M$ and $R$, and (3)~significant bias in $M$ and $R$.  Thus,
at least for the systematic differences from the model explored by
\cite{2013ApJ...776...19L}, if the fit is good and the
constraints are strong, the inferred values of the mass and
radius are reliable.

Valuable extra information could be obtained from the identification
of atomic lines from the surfaces of rotating neutron stars.  No such
line has been confirmed, and indeed even if a line-like feature is seen in a spectrum it is not trivial to identify the $z=0$ atomic transition corresponding to the line.  One such identification was claimed from an analysis of stacked
bursts from EXO~0748 \cite{2002Natur.420...51C}, but an additional
long look at the star found it in another state that had no lines
at all, whether zero redshift or from the surface, and thus was
unable to confirm the lines \cite{2008ApJ...672..504C}. The spin
frequency of this star is 552~Hz \cite{2010ApJ...711L.148G} rather
than the originally claimed 45~Hz \cite{2004ApJ...614L.121V}, and hence
one might expect that Doppler smearing would make a sharp line
undetectable (although  note that \cite{2013ApJ...766...87B} suggest
that sharp lines would still be visible; if this result is confirmed,
it means that there are better prospects for sharp lines than
previously thought).   If future large-area instruments are able to
not only detect such features but also measure them precisely, then both the redshift from the surface and the
linear speed of the surface at the spot, as well as possibly even
frame-dragging effects, could be inferred
\cite{2006ApJ...644.1085B}.  This would allow many degeneracies to be
broken and would lead to much more precise constraints on neutron
star masses and radii (and moments of inertia from frame-dragging). 
Note that such measurements will only be possible from actively
accreting stars, because heavy elements sink in the atmospheres of isolated neutron stars within seconds \cite{1980ApJ...235..534A}. It has also been
proposed that the equivalent width of the line will allow a
measurement of the surface gravity, and hence that $M/R$ (from the
redshift) and $M/R^2$ (from the surface gravity) can be measured independently (e.g.,
\cite{2005ApJ...629..998C}).  In principle this is also possible using
a non-thermal continuum spectrum, but this would require exceptional
data.

\subsection{Maximum spin rate}

Another method that has been suggested to constrain the radius
(or more properly, the average density) is measurements of spin
frequencies: a high enough spin frequency from any star would
rule out the hardest equations of state.  Unfortunately, no
confirmed spin frequency is high enough to place significant
limits on dense matter.  The maximum spin frequency for a star
of mass $M$ and radius $R$ is roughly \cite{1994ApJ...424..823C}
\begin{equation}
\nu_{\rm max}=1250~{\rm Hz}\left(M/M_\odot\right)^{1/2}
(R/10~{\rm km})^{-3/2}\; .
\end{equation}
This is $\approx 800$~Hz for $M=1.4~M_\odot$ and $R=15$~km.  In comparison,
the highest frequency ever established is 716~Hz, for PSR~J1748--2446ad
in the globular cluster Terzan 5 \cite{2006Sci...311.1901H}. 
Weak evidence for an 1122~Hz signal during thermonuclear bursts from
XTE J1739--285 has been claimed \cite{2007ApJ...657L..97K} 
but not confirmed.  Even if it were
confirmed there would have to be strong evidence that this was the
fundamental frequency instead of the first overtone; the overtone
can be dominant, as was shown for IGR J17511-3057 
\cite{2010arXiv1005.5299A}.

Given that neutron stars could spin faster than they do, and that X-ray
observations using the {\it Rossi} X-ray Timing Explorer are not biased
against signals with $\nu>1000$~Hz \cite{2008AIPC.1068...67C}, why do
the stars not spin faster?  A conservative answer that is in agreement
with all data is that magnetic torques during accretion and spinup limit
the frequency.  The required dipolar field strengths $\sim 10^8$~G agree
with the fields inferred from the spindown of their descendants (the
rotation-powered millisecond pulsars), and in particular with the
adherence of those pulsars to the spin-up line
\cite{2009ASSL..357...19M}.  Fields of this strength are also consistent with the spin behavior of stars between transient outbursts \cite{2011ApJ...738L..14H,2012ApJ...746....9P}.  Another possibility is that in many cases
not enough matter has been accreted to reach maximum spin. An exciting
longshot that has received much attention due to the rapid improvements
in ground-based gravitational wave detectors is that nonaxisymmetries in
neutron stars, e.g.,  Rossby waves \cite{1999ApJ...510..846A,
2013arXiv1308.3685C,2013arXiv1305.2335B} (see \cite{2012MNRAS.424...93H,
2013ApJ...773..140M} for recent observational constraints) or perhaps
accretion-induced lumpiness in the stars \cite{2000MNRAS.319..902U}, might
counteract accretion spinup via emission of gravitational radiation.  One
way to test this hypothesis is to observe systems that have had multiple
transient episodes, because the spindown between active phases could 
indicate whether gravitational radiation (which would depend only on the
long-term average accretion rate rather than the instantaneous rate) 
emits angular momentum at the required rate.  Two such systems
(SAX~J1808--3658 and IGR~J00291+5934) have been observed with the required precision.
In both cases there is no evidence that gravitational radiation induced
spindown is occurring, but within the observational uncertainties there
is room for contributions at the tens of percent level \cite{2009ApJ...702.1673H,2010arXiv1006.1303P}.
Whatever the reason for the spin ceiling, at this stage there are no
known neutron stars with spin frequencies high enough to rule out any
plausible equation of state.

\subsection{Kilohertz QPOs}

Kilohertz quasi-periodic brightness oscillations (kHz QPOs) from accreting
neutron stars have been proposed to constrain the masses and
radii of the stars.  The basic phenomenology of kHz QPOs is that there are
commonly two relatively narrow ($Q\equiv \nu/{\rm FWHM}\sim
20-200$) QPOs that appear in the power density spectra of
more than 25 neutron star low-mass X-ray binaries.  Both
frequencies vary by hundreds of Hertz between and during
observations. The higher-frequency of the two often reaches
$\nu>1000$~Hz, and the lower-frequency peak (which is often the
sharper one, and also commonly has a larger fractional root mean
square amplitude) has a frequency less than that of the upper peak
by a characteristic but not exactly constant 
amount that may be related to the spin
frequency of the star (but see \cite{2007MNRAS.381..790M} for a
dissenting opinion).  

Most, but not all, modelers identify the
upper peak frequency with an orbital frequency at some
characteristic radius around the star.  If this is true, it means
that the star must fit inside that radius, as must the radius of
the innermost stable circular orbit (ISCO) predicted by general
relativity; the latter condition follows because matter inside the
ISCO will fall rapidly towards the star and thus prevent it from forming
high quality factor oscillations. As derived by \cite{1998ApJ...508..791M},
for a neutron star with a dimensionless rotation parameter $j\equiv
cJ/GM^2$ these conditions limit the mass and radius to $M_{\rm max}=2.2~M_\odot 
(1+0.75j)(1000~{\rm Hz}/\nu_{\rm QPO})$
and $R_{\rm max}=19.5~{\rm km}(1+0.2j)(1000~{\rm Hz}/\nu_{\rm QPO})$
for an upper peak frequency $\nu_{\rm QPO}$.  The highest confirmed
QPO frequencies are all less than 1300~Hz (see, e.g., 
\cite{2009MNRAS.399.1901B} for a discussion of the 1330~Hz QPO once
suggested for 4U~0614+09), so at this stage the constraints are
not restrictive.  If broad iron lines from the inner disk are
discovered simultaneously with kilohertz QPOs, this will provide
another measure of the mass because the line breadth gives
$\sqrt{M/r}$ whereas the QPO frequency gives $\sqrt{M/r^3}$ at the
orbital radius $r$ (see \cite{2008ApJ...674..415C} for current
data and \cite{2011MNRAS.415.3247B} for future prospects; note
that the Kerr spacetime is not an adequate approximation for sufficiently rapidly
rotating neutron stars \cite{1998ApJ...509..793M}).  Similarly, if
reverberation mapping can establish an absolute time scale for the
system, this might yield masses and radii
\cite{2013ApJ...770....9B}.

If the orbital frequency of the ISCO is established for a star, then the
mass of the star is known to within a small uncertainty related to the star's dimensionless angular momentum parameter.  After doubt was cast on initial
claims of ISCO signatures \cite{1998ApJ...500L.171Z} because of the
complex relation between count rate and QPO frequency in these stars
\cite{1999ApJ...511L..49M}, recent analysis of the RXTE database
has suggested that in many stars the predicted sharp drop in
quality factor and gradual drop in amplitude \cite{1998ApJ...508..791M}
are seen at a frequency that is consistent across a wide range in
count rate and X-ray colors \cite{2005MNRAS.361..855B,
2005AN....326..808B,2006MNRAS.370.1140B,2007MNRAS.376.1139B}.
The independence of this behavior from proxies of mass accretion
rate such as count rate and colors led these authors to suggest
that a spacetime marker such as the ISCO was the most likely reason
for the observed phenomena.  If so, this represents a confirmation
of a key prediction of strong-gravity general relativity (the ISCO),
and implies masses greater than $2.0~M_\odot$ for some neutron stars,
which would be highly constraining on equations of state.  

Such important implications demand careful examination.  For example,
\cite{2006MNRAS.371.1925M} notes that the maximum quality factor
achieved by neutron star LMXBs, versus their average luminosity, has a
shape similar to the quality factor versus radius in individual stars,
and uses this to conclude that other factors operate and that the drop
in quality factor might not be caused by approach to the ISCO.   Indeed,
other factors do influence the quality factor; for example, the
high-luminosity stars plotted by \cite{2006MNRAS.371.1925M} have
geometrically thick disks that cannot produce high-$Q$ oscillations. 
This is thus not directly relevant to the arguments made by
\cite{2005MNRAS.361..855B,
2005AN....326..808B,2006MNRAS.370.1140B,2007MNRAS.376.1139B} because the
stars they examined are all low-luminosity, and hence the results of
\cite{2006MNRAS.371.1925M} do not address whether the ISCO causes the
behavior observed in those stars.  Nonetheless, the complex
phenomenology of kHz QPOs and the lack of first-principles numerical
simulations that display them means that claims that the ISCO has been
detected must be treated with care.

\subsection{Other methods to determine the radius and future prospects}

There are two other noteworthy ways to measure the radius that have been suggested
with particular applications to the double pulsar J0737.  The
first involves the binding energy of the lower-mass Pulsar B in
that system.  Based on an idea originally proposed by
\cite{1984ApJ...277..791N}, \cite{2005MNRAS.361.1243P} suggested
that this neutron star formed via an electron-capture supernova
(in which electron captures onto Mg and then Ne in a core cause
a loss of pressure support) rather than the usually considered
collapse of an iron core when it goes above the Chandrasekhar
mass.  As this electron capture happens at a very specific central
density of $4.5\times 10^9$~g~cm$^{-3}$ that corresponds to a
well-defined baryonic mass of $M_{\rm bary}=1.366-1.375~M_\odot$
\cite{2005MNRAS.361.1243P}, if one could identify this baryonic
mass with the gravitational mass $M_{\rm grav}=1.2489\pm 0.0007~M_\odot$
of Pulsar B then one would have an extremely precise constraint
that would suggest a fairly hard equation of state.  However,
there is an unknown amount of fallback after the initial collapse,
so the electron capture scenario only suggests that 
$M_{\rm bary}(M_{\rm grav}=1.2489~M_\odot)>1.366-1.375~M_\odot$.
This allows a wide variety of equations of state.

The second method stems from the observation that spin-orbit coupling, which is dominated in this
system by the 23~ms period Pulsar A instead of the 2.8~s period Pulsar
B, leads to precession of the orbital plane and additional pericenter
precession \cite{1975PhRvD..12..329B,
1988NCimB.101..127D,2004Sci...303.1153L,2005ApJ...629..979L,
2008ARA&A..46..541K}.  Orbital plane precession will be difficult to
measure, but the required precision for the extra pericenter
precession could be reached in the next few years
\cite{2005ApJ...629..979L}. Such a measurement would effectively
determine the moment of inertia $I$ of Pulsar A (potentially to 10\%;
\cite{2005ApJ...629..979L}), and given that $I\sim MR^2$ and $M$ is
known well, this would amount to a $\sim 5$\% measurement of the
radius.  It is unclear whether this will be reachable in practice,
because of the confusing effect of the unknown acceleration of the
center of mass of the binary within the Galactic potential
\cite{2008ARA&A..46..541K}.

In general, the methods discussed in this section have uncertainties that 
are dominated by systematics.  As a result, future progress depends on
both better observations and better models.  For example, there is
a recent set of magnetohydrodynamic simulations that aim to reproduce
kHz QPO phenomenology in accreting weakly magnetic neutron stars
\cite{2003ApJ...595.1009R,2004ApJ...610..920R,2005ApJ...633..349K,
2006AdSpR..38.2887R,2007MNRAS.374..436L,2007ApJ...670L..13L,
2008ApJ...673L.171R,2009MNRAS.398..701K,2013arXiv1309.4383B}.  These
simulations may ultimately result in solid understanding of kHz QPOs that
could be used to interpret observations.  The simulations are
extremely challenging, however, and may take many years to get to
the point of full reliability.  Some of the apparently disparate
constraints may eventually be coupled through the recently
discovered ``I-Love-Q" relations between the moment of inertia
I, Love number, and quadrupole moment Q \cite{2013PhRvD..88b3009Y}, 
although the important issues of systematics must be solved first.

\section{Cooling of Neutron Stars}
\label{sec:cooling}

As noted in \S~\ref{sec:nuclear}, cooling processes are
sensitive to different aspects of the equation of state than
are the maximum mass and the mass-radius relation.  This in
principle means that observations of cooling neutron stars can
give us a complementary tool with which to constrain the
properties of dense matter.  In fact, current data are broadly
consistent with the dense matter in neutron stars not having significant
contributions from exotic phases (modulo some complications
we shall discuss), but unfortunately as we will see this is
a very blunt tool and there is plenty of room for exotic
matter.

X-rays from cooling neutron stars were originally proposed in
the mid-1960s \cite{1964PhRvL..12..413C} 
as one of the few ways that these stars
could be detected.  In this section we will focus on cooling
theory and observations that bear on the matter at the cores of neutron
stars.  We will thus not discuss, e.g., the cooling of
transiently accreting neutron stars (see \cite{2004ARA&A..42..169Y,
2006NuPhA.777..497P} for recent
reviews) that return to quiescence after a years-long outburst
has raised the crust out of thermal equilibrium with the
core, because their cooling curves depend primarily on
processes in the crust.
 
Broadly speaking, after a neutron star forms in a
supernova (where at birth its temperature is roughly the
virial temperature $T_{\rm vir}\sim GMm_n/(Rk)\sim 10^{12}$~K),
the star goes through a phase of duration $\sim 10^{4-6}$~yr
in which its cooling is dominated by neutrino losses from the
core.  After this point the star cools mainly by photon
luminosity from the surface, where the energy from the core
is transported conductively until the density is low enough
that radiative processes take over (this typically occurs in
the outer crust).  The temperature of a neutron star at a given
age therefore depends strongly on how long neutrino emission
dominates.  The temperature tool is blunt because there are many processes that
can enhance neutrino production, and many processes that can
suppress it, and thus mere measurement of the temperature with age
of neutron stars would not allow us to distinguish easily
between the multiple candidate effects.

\subsection{The URCA processes}

To start, we note that as the star cools down from its 
high-temperature birth the processes $p+e^-\rightarrow n+\nu_e$
and $n\rightarrow p+e^-+{\bar\nu}_e$ can produce neutrinos
efficiently.  In this context these are known as the URCA
processes, named thus by George Gamow after the Urca casino
in Rio de Janeiro because the URCA processes are a perfect
sink for energy just as the casino is a perfect sink for money!
Phase space considerations indicate that the URCA emissivity
(energy per volume per time) scales as $T^6$.

As the temperature cools below the Fermi temperature, however,
we can see that these processes become impossible unless the
proton to neutron ratio is sufficiently high.  The derivation
of the critical ratio is similar to what we presented in
\S~\ref{sec:nuclear}: the momenta of
the particles, which are dominated by Fermi momenta, must
satisfy the triangle inequality $p_n\leq p_p+p_e$.  The Fermi
momenta of all three particles
are determined by their respective number densities, so
this means $n_n^{1/3}\leq n_p^{1/3}+n_e^{1/3}$.  Charge
neutrality means $n_e=n_p$, so $n_n^{1/3}\leq 2n_p^{1/3}$
and thus the criterion for the URCA process to be possible
is $n_n\leq 8n_p$.  This is on the low
side for many traditional equations of state but can be achieved
in some cases at high density.  Note that if muons are also
present (these are higher-mass analogs to electrons with the same electric charge),
and thus electrons have a number density $n_e=xn_p$ with $x\leq 1$, the
criterion becomes $n_n\leq (1+x^{1/3})^3n_p$, so even higher
proton fractions would be required.  

If the URCA process is suppressed, bystander particles can
soak up the extra momentum, e.g., $n+n\rightarrow n+p+e^-+{\bar\nu}_e$.
This is usually called the modified URCA process to distinguish
it from the direct URCA (sometimes DURCA) processes described
above.  Given that the neutrons are degenerate, only a fraction
$T/T_F$ of them can interact.  Thus the modified URCA process
is suppressed by a factor $(T/T_F)^2$ (one factor is for the initial
neutron and one is for the final) compared to the direct URCA process.
Given that in a neutron star core $T_F\sim 10^{12}$~K and $T$
can be $\sim 10^9$~K, the suppression factor can easily be a million. 

The huge difference between the direct and modified URCA rates
means that if enough protons are present, or if there are any
other channels for neutrino production, then cooling can be enhanced
dramatically.  We now consider such channels.

\subsection{Additional neutrino production channels and suppression}

As described in the lucid review of \cite{2006NuPhA.777..497P}, 
hyperons can produce
neutrinos via, e.g., $\Sigma^-\rightarrow\Lambda+e^-+
{\bar\nu}_e$, $\Lambda+e^-\rightarrow\Sigma^-+\nu_e$, and
processes that involve both hyperons and nucleons.  If
condensates form (the leading candidates are of pions or kaons)
then the condensate acts as an effectively infinite reservoir
of momentum.  This produces channels such as
$n+\langle\pi^-\rangle\rightarrow n+e^-+{\bar\nu}_e$ and
$n+\langle K^-\rangle\rightarrow n+e^-+{\bar\nu}_e$, where
the angle brackets indicate the condensate.  The kaon process
involves a strangeness change and is thus less efficient than
the pion process, modulo effects related 
to the medium \cite{2006NuPhA.777..497P}.
If deconfined quarks are a significant degree of freedom then
other direct URCA-like processes can emerge, such as
$u+e^-\rightarrow d+\nu_e$ and its inverse ($e^+$ could be
present in negatively charged quark matter, but the processes
are the same), and 
$u+e^-\rightarrow s+\nu_e$ and its inverse (suppressed by an
order of magnitude because of the strangeness change).
All of these processes are orders of magnitude more efficient
than modified URCA, and all scale as $T^6$.

A separate channel has the interesting property that it can
both increase and suppress cooling, depending on the details.
This is the transition to a superfluid state.  This transition
occurs via Cooper pairing of the neutrons, and the immediate
effect is the emission of a neutrino-antineutrino pair:
$n+n\rightarrow [nn]+\nu+{\bar\nu}$.  Given that one fewer
effective particle is involved than in modified URCA the
emissivity scales as $T^7$ instead of $T^8$, so at
$T\sim 10^9$~K the emissivity of Cooper pairing can be
comparable to or greater than modified URCA.  As the shell
where the temperature is less than the superfluid critical temperature
moves inwards, this can therefore enhance neutrino emission.

In the long term, however, the pairing produces an energy
gap $\Delta\sim kT_c$ (where $T_c$ is the superfluid critical
temperature) at the Fermi surface that suppresses
processes by a factor of order $e^{-\Delta/kT}$ for 
$T\ll T_c$, modulo details of the phase space and
the temperature dependence of $\Delta$.  The suppression can
thus be extremely dramatic.  It occurs for both the
neutrino emissivity and for the specific heat (with different
factors).  See, e.g., Figure~5 of \cite{2006NuPhA.777..497P}
for plots of some of the suppression factors (called
control functions there).  The critical temperature and
energy gap are extremely difficult to calculate from first
principles but most estimates suggest $\Delta\sim 1$~MeV, 
which corresponds to $T\sim 10^{10}$~K and is thus enough
larger than core temperatures to make a significant difference.

Cooper pairing can also occur in deconfined quark matter.
In that context it is much more complicated than in ordinary
matter because quarks have different colors, flavors, and
masses.  Multiple types of condensation are therefore
possible.  The color gap is estimated to be 
$\sim 50-100$~MeV \cite{2006NuPhA.777..497P}, which is huge compared to 
internal temperatures and is thus potentially quite important.

There are also neutrino emission processes in the crust
of the star, such as plasmon decay and neutrino pair
bremsstrahlung from electron-ion and electron-electron
interactions.  These could contribute for middle-aged
stars, when the core processes are suppressed by the
superfluidity gap and the crust is warm enough to produce
neutrinos. 

\subsection{Photon luminosity and the minimal cooling model}

When neutrino emission has tapered off, further cooling of
the star is controlled by conductive transport in the dense
layers of the star (where electrons are degenerate and thus
have large mean free paths) and by radiative transport 
nearer the surface.  There is a minimum in the overall efficiency of
energy transport at the ``sensitivity layer" where conduction
hands off to radiation (where electrons are only partially
degenerate).  This thus acts as a bottleneck and largely
determines the overall cooling rate; the density at the
sensitivity layer increases with increasing temperature.
For temperatures in the observed range $\sim 10^{5-6}$~K the sensitivity 
layer is at a density $<10^9$~g~cm$^{-3}$, where pycnonuclear
fusion is not guaranteed to convert light elements to heavy
ones.  The electron thermal conductivity in liquid ions
depends on their charge $Z$ as $\sim 1/Z$ \cite{2006NuPhA.777..497P}, 
so differences
in the composition of the upper layers (due, e.g., to different
fall-back after core collapse) could have an influence on the
thermal evolution of neutron stars.

Conduction dominates at high enough densities that magnetic fields are
likely to be unimportant: at typical densities $\rho\gta 10^6~{\rm
g}~{\rm cm}^{-3}$ the Fermi energy is $E_F\gta m_ec^2$, implying that
the magnetic field needs to be $B\gta B_c=m_ec^3/(\hbar e)=4.414\times
10^{13}$~G to have a significant influence.  However, more moderate
magnetic fields can affect radiative transport near the surface. More
specifically, suppression of electron motion across field lines yields
anisotropic conduction.  This can produce strong anisotropies in the
emergent radiation, but the overall effect on cooling is relatively
small \cite{2006NuPhA.777..497P}.   Ultimately, the radiation emerges
with some spectrum and an effective temperature $T_{\rm eff}$ that can
be defined as
\begin{equation}
T_{\rm eff}^4\equiv {1\over{4\pi}}\int \int T_s^4(\theta,\phi)\sin\theta
d\theta~d\phi\; .
\end{equation}
Here, the
local effective temperature $T_s(\theta,\phi)$ at each location $(\theta,\phi)$ on the surface is defined
via $\sigma_{\rm SB} T_s^4(\theta,\phi)=F(\theta,\phi)$ where $F$
is the emergent photon flux and these quantities are
appropriately redshifted.  If the emergent spectrum is
close to a blackbody then one can estimate the temperature
without knowing the distance to the star (see \S~\ref{sec:radius}),
but as we discussed earlier atmospheric effects cause the spectrum to
deviate from a blackbody form, which complicates inferences.

With this physics in place one can construct a surface temperature
versus age curve by including a specific choice for the uncertain
core processes as well as choosing the composition of the surface
layers.  A particularly useful choice is the ``minimal cooling
model" (see \cite{2009ApJ...707.1131P} for a recent treatment), in which
one assumes no exotic components or direct URCA but does include
Cooper pairing, crustal bremsstrahlung, and all relevant photon
processes.  This represents the smallest amount of cooling that is
realistic.  As a result, if there is evidence that at least some
neutron stars are significantly cooler than they are predicted to be in this
model, that might suggest exotic components (although it does not
work the other way around; conformity with minimal cooling might
mean the suppression effects are important).  To see how this
cooling model fares we now turn to observations.

\subsection{Observations and systematic errors}

Before discussing the observations we must issue a series
of caveats.  There are many reasons why it is difficult to generate reliable points
on a temperature-age curve, including
the interpretation of the spectrum, other possible heating
sources, and challenges with age estimation.  We now
discuss these in order before finally evaluating the best
current data on cooling neutron stars.

A neutron star with a surface effective temperature of
$10^6$~K and a radius of 10~km has a luminosity of
$L\sim 10^{32}$~erg~s$^{-1}$, which at a distance of
3~kpc gives a detector flux of $F\approx 7\times
10^{-13}$~erg~cm$^{-2}$~s$^{-1}$ and corresponds to
$\sim 2$ counts per second for a 1000~cm$^2$ detector.
It is therefore possible to get a reasonable number of
photons over a long observation, although the thermal
peak of $\sim 0.2~{\rm keV} (T/10^6~{\rm K})$ is strongly
susceptible to interstellar absorption.  

There are also complications with the atmospheric model, as we
discussed when we examined radius estimates for cooling neutron
stars.  For example, compared to a blackbody with the same
effective temperature, an unmagnetized hydrogen atmosphere has a
strong excess at higher energies when absorption dominates the opacity,
because these opacities
scale as $\nu^{-3}$ (e.g., \cite{1986rpa..book.....R}).  
Magnetized hydrogen has less of an excess because strong fields
increase the binding energy of atoms (e.g., Problem 3 in \S112 of
\cite{1965qume.book.....L}), and heavier elements also have more
bound electrons and thus less of an opacity deficit at high
energies compared to lighter elements
\cite{1987ApJ...313..718R,1991MNRAS.253..107M,1992MNRAS.255..129M,
1997ApJ...479..347R}.  Observational support for the diffusive
burning of hydrogen or helium
\cite{2003ApJ...585..464C,2010ApJ...723..719C} may have been
obtained from the evidence that the atmosphere of Cas A is
dominated by carbon \cite{2009Natur.462...71H}. As a result, unless there
is a clear statistical preference for one atmospheric model versus
another (which is not currently the case for any star), the
temperature will be uncertain.

We must also be cautious because in addition to simple
cooling there are various heat sources that could
contribute.  These include magnetic dissipation (likely
only important for highly magnetic neutron stars) and
magnetospheric emission.  The latter is likely to produce
a nonthermal spectrum, which is indeed seen in some
stars.  One could argue with some justice that if we
are looking for cases where the temperature is {\it less}
than predicted by the minimal cooling model, other heat
sources will only mean that such evidence is even stronger.
The tricky part comes when one subtracts off nonthermal
components to estimate the underlying thermal emission;
if the other components were oversubtracted this would
lead to an underestimate of the cooling temperature.

The final caveat relates to the age.  If the star was
born with a much more rapid spin than it has now
and it has slowed down exclusively by magnetic dipole
radiation then its current age is $P/(2{\dot P})$,
where $P$ is the current spin period and ${\dot P}$
is the current spin derivative.  In some cases this age
estimate can be checked using a kinematic age, either
from a supernova remnant or from the angular distance
above the Galactic plane (where massive stars are born)
divided by the angular proper motion away from the plane.
Unfortunately there is often a discrepancy between these
estimates of a factor of three or more, so all of these
numbers must be treated with caution.

With the preceding in mind, Figure~8 in \cite{2009ApJ...707.1131P} shows
the comparison between the minimal cooling model (with 
light or heavy element envelopes) and the data.  Within
the significant uncertainties we note that all of the
data are consistent with the minimal cooling model, although
there is some evidence that variation in
the envelope composition is needed to explain the data.
As \cite{1992ApJ...394L..17P} note, however, there is a
significant selection effect at work: if there are stars
that have cooled rapidly, they are obviously more difficult
to see.  It is thus possible that using our current
satellites we can only observe comparatively hot stars.

Recently it was suggested that the neutron star in the supernova
remnant Cas A has cooled very rapidly over the past decade or
so \cite{2009Natur.462...71H}.  Further observational analyses,
especially taking into account the complexity of the surrounding
supernova remnant emission \cite{2013arXiv1306.3387E} and the
possibility of changes in the calibration, emitting region size,
or absorbing column \cite{2013arXiv1311.0888P} have made it much less clear
that there actually is anomalously fast cooling.  If the evidence
strengthens for such cooling, it has potentially exciting
implications for the physics of the interior of this neutron
star, with the leading idea being Cooper pair creation in the 
superfluid \cite{2011MNRAS.412L.108S,2011PhRvL.106h1101P}.
One-pion exchange and polarization effects could also play a
role \cite{2013arXiv1308.4093B}.

\subsection{Current status and future prospects}

Currently there is no evidence that exotic components are
{\it necessary}, although given the large uncertainties in
data they could certainly be {\it accommodated}.  The possible
lack of exotic components is consistent with the tentative
evidence presented in \S~\ref{sec:radius} that some
neutron stars have masses $\gta 2.0~M_\odot$, and could
mean that nucleonic degrees of freedom dominate the internal
structure of neutron stars.  However, the data are not clear.

To improve the observational situation it will be necessary to
have much larger-area future X-ray observatories, such as the
planned ATHENA+ \cite{2013arXiv1306.2307N}.  The resulting
high-signal observations would play two important roles: 
(1)~they would reduce the bias against rapidly cooled stars,
and would thus possibly reveal evidence for exotic components,
and (2)~they would allow us to distinguish empirically between
different candidate atmospheric spectra (nonmagnetic hydrogen
or helium, magnetic atmospheres, and even the straw man blackbody
spectrum), and thus have greater confidence in the inferred
temperatures.  It will be more challenging to come up with
much more precise and accurate ages, but it could be that future
radio arrays such as the Square Kilometer Array will allow us
to determine the proper motion and kinetic ages more accurately.

\section{Gravitational Waves from Coalescing Binaries}
\label{sec:future}

In the next few years it is expected that the worldwide gravitational
wave detector network will achieve sufficient sensitivity to detect
$\sim 0.4-400$ NS-NS mergers per year, and an uncertain
number of NS-BH coalescences (see \cite{2010CQGra..27q3001A} 
for a recent discussion of the
predicted rates and their substantial uncertainties).
In this section we summarize what can be learned from such observations.
It has also been proposed that observation of neutron star oscillation
modes stimulated by mergers or (at much lower amplitude) glitches
would yield important
information about the stars (see, e.g., \cite{2008CQGra..25v5020V,
2012IJMPS..18...48C,2012PhRvD..86f3001B,
2013PhRvD..87d4043C,2013PhRvD..88d4052D});
this is true, but the amplitudes are likely to be significantly below
the amplitudes of the coalescence signal, so we will not focus
on oscillations here.

In brief, for a binary of masses $m_1$ and $m_2$ and thus
total mass $M=m_1+m_2$ and symmetric mass ratio 
$\eta=m_1m_2/M^2$ (note that $\eta\leq 0.25$, with the maximum
occurring for $m_1=m_2$), 
the combination $\eta M^{5/3}$ will be determined with high
precision and could lead to significant constraints if a high enough
mass binary is detected.  The individual masses and the average
density of the neutron stars will be more challenging to measure, but
they seem within reach for the strongest events.

In more detail, we note that to lowest order gravitational radiation changes the binary orbital frequency at a rate
(see \cite{1964PhRv..136.1224P} for the rate of change of the semimajor axis)
\begin{equation}
{df\over{dt}}={96\over 5}(4\pi^2)^{4/3}G^{5/3}\eta M^{5/3}f^{11/3}c^{-5}
(1-e^2)^{-7/2}(1+73e^2/24+37e^4/96)\; .
\label{eq:Peters}
\end{equation}

Now consider a NS-NS binary.  When the orbital separation is much
greater than the radii of the neutron stars, the inspiral of the
binary proceeds almost as it would if the stars were point masses.
From equation~\ref{eq:Peters} we see that the mass combination
$M_{\rm ch}^{5/3}\equiv\eta M^{5/3}$ (where $M_{\rm ch}$ is called
the ``chirp mass") determines the frequency and amplitude evolution.  It can
thus be measured with precisions better than 0.1\% in many cases
\cite{2009CQGra..26t4010V}.  
Even with no additional information we can obtain a strong
lower limit to the total mass $M$ by setting
$\eta=0.25$, and thus can obtain a strong lower limit $M/2$ to the greater
of the two masses.  It could be that if tens to hundreds of NS-NS mergers are
observed per year, and if these can be distinguished clearly from NS-BH and BH-BH mergers, then a small number of them will have $M>4.0~M_\odot$
and thus it will be possible to establish rigorously that
$M_{\rm max}>2.0~M_\odot$.

We might not be this lucky.  If the NS-NS
detection rates are as low as $\sim 1$~yr$^{-1}$
then high-mass binaries might not be sampled.  If the rate is
tens per year but the chirp masses do not imply a high minimum
mass it could be that the upper limit to neutron star masses is close to the $\sim 2~M_\odot$ that we have already established from electromagnetic observations, but it could also
be that there is essentially only one way to form NS-NS systems and thus that those systems will tend to have similar chirp masses.  

\begin{figure}[t]
\includegraphics[scale=0.6]{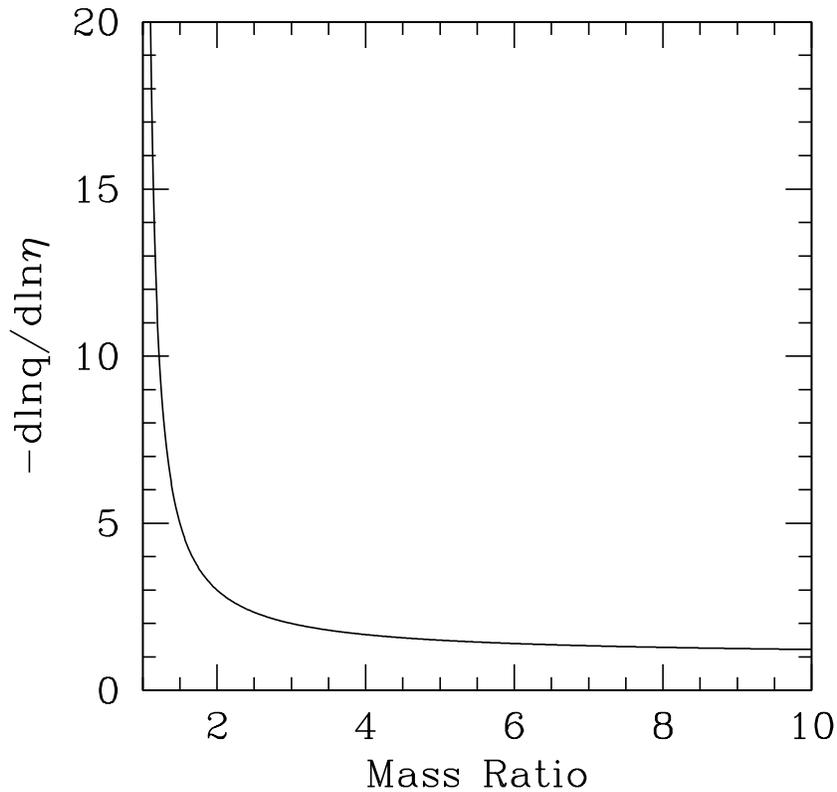}
\caption{Logarithmic derivative of the mass ratio $q\equiv m_1/m_2\geq 1$
with respect to the symmetric mass ratio
$\eta$.  For double neutron star binaries, in which $q\sim 1$,
even a small error in $\eta$ leads to significant uncertainty
in $q$.  As a result, gravitational waves from
double neutron star systems are not ideal
for precise mass measurements of individual neutron stars.}
\label{fig:dlnq}
\end{figure}

In these cases more information is needed.  When the inspiral is
followed to higher post-Newtonian order the additional terms,
which involve tidal effects, have
different dependences on $\eta$ and $M$ than does the lowest-order
expression, so this can be used to break the
degeneracy and infer the two masses separately.  For systems in
which the two masses are comparable (as in a NS-NS system) this
requires very high precision measurements of $\eta$.  Note, for
instance, that whereas $\eta=0.25$ implies a 1:1 mass ratio,
$\eta=0.24$ implies 1.5:1.  Figure~\ref{fig:dlnq} shows that for nearly equal masses
a very small fractional error in $\eta$ can still imply a
large fractional error in the mass ratio $q\equiv m_1/m_2\geq 1$.  
The sensitivity is naturally less
away from the maximum, which might lead one to suppose that
NS-BH binaries, which are asymmetric in mass, would provide
greater prospects for NS mass measurements.  For example, suppose
that $M_{\rm ch}=2.994~M_\odot$ is measured with effectively zero
uncertainty, and that $\eta$ is constrained to be between 0.1 and 0.11.
Then the lighter component of the binary has a mass between
$1.34~M_\odot$ and $1.42~M_\odot$ and is thus well known
despite the significant fractional uncertainty in $\eta$.

The tradeoff is that black holes may typically have large enough masses that
neutron stars spiral into them without significant tidal effects
\cite{2005ApJ...626L..41M}, although higher harmonics
can still be important and as partial compensation they will
appear at frequencies of greater sensitivity in ground-based
detectors than will the corresponding harmonics in NS-NS
systems.  In addition black holes, unlike the
neutron stars in NS-NS binaries, may well have large enough spins
to affect the last part of the inspiral and thus compromise
parameter estimation due to the greater complexity of the waveforms. 
No pulsar in a NS-NS binary has a spin period shorter than 23~ms 
\cite{2003LRR.....6....5S}, so the spin parameters of these neutron stars are
$j\equiv cJ/GM^2\lta 0.02$.  In contrast, although inferences of
the spins of stellar-mass black holes are not yet fully vetted,
there is growing evidence that many of them have spin parameters
of several tenths, with some possibly approaching the black hole
maximum $j=1$ \cite{2007ARA&A..45..441M,2010AIPC.1248..101M}.  The likely
evolutionary difference is that in the supernova that creates a
neutron star or black hole, much greater amounts of mass fall back
to create a black hole than a neutron star, and this mass will
come from significant radii that thus carry considerable angular
momentum.  Accretion from a binary subsequent to the production of
a black hole has little effect on either the mass or the angular
momentum of the hole (see, e.g.,  \cite{1999MNRAS.305..654K}). 
Neutron stars in the high-mass X-ray binaries that could create a
double neutron star pair accrete little mass in the active phase
and, empirically, seem to have relatively strong magnetic fields,
so accretion does not spin them up to high rotation rates. 

These statements are not absolute.  It could be that there are
slowly-spinning black holes in BH-NS binaries, or rapidly spinning
neutron stars in NS-NS binaries (especially if they are produced
dynamically in globular clusters).  However, the expectation at this
time is that until tidal effects become important NS-NS waveforms
will be easier to interpret than those from BH-NS coalescences.

\begin{figure}[t]
\includegraphics[scale=0.55]{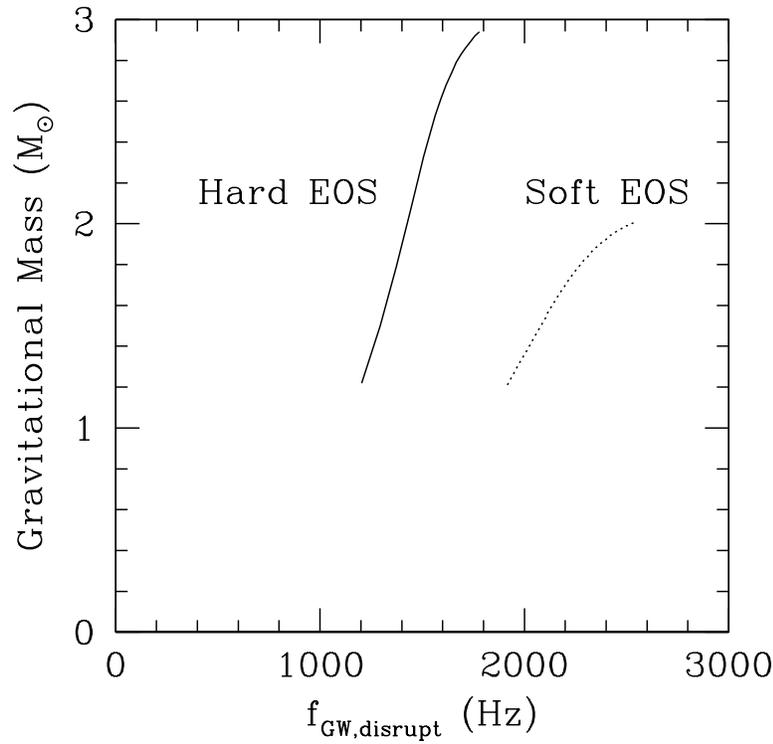}
\caption{Gravitational wave frequency at the point of tidal
disruption as a function of neutron star mass.  The solid and dotted lines refer, respectively, to the hard and soft equations of state from \cite{2013ApJ...773...11H}.  In both cases we assume that the companion is a
neutron star of equal mass.  Tidal effects will affect
the phasing of the inspiral below $f_{\rm GW,disrupt}$, but the
effects fall off rapidly with decreasing frequency.  Good
sensitivity at high frequencies is essential for these effects to
be detected.}
\label{fig:tidal}
\end{figure}

This leads us to what those tidal effects might tell us about neutron
stars.  At large separations compared to the neutron star radii,
nonlinear mode couplings due to tides will not affect the inspiral phase 
significantly \cite{2013arXiv1307.2890V}.  Most of the information
that can be obtained from tidal phase deviations from point-mass
inspiral exists at high frequencies \cite{2010PhRvD..81l3016H}, and
recent comparisons of analytical theory with numerical simulations
suggest that with some calibration the theory does extremely well
\cite{2012PhRvD..85l3007D,2012PhRvD..86d4030B,2012PhRvD..85d4045F,
2013PhRvD..87d4001H,2013arXiv1303.6298L}.  An explicit comparison
of the expected signals from piecewise-polytropic equations of state
suggests that a difference of only 1.3 km in radius could be
distinguished for NS-NS systems out to 300~Mpc with optimal direction and binary orientation (this corresponds roughly to a direction- and orientation-averaged distance of 140~Mpc) given a full hybrid post-Newtonian plus numerical relativity waveform \cite{2013PhRvD..88d4042R}.  See Figure~\ref{fig:tidal} for an indication of the different frequencies of tidal disruptions implied by two candidate equations of state.

Inspirals of neutron stars into black holes have not yet been
examined using as much care with respect to phase deviations.
Recent work suggests that although if the mass ratio is as high
as 6:1 a NS-BH coalescence will be indistinguishable from a BH-BH
coalescence with the same masses \cite{2013PhRvD..88f4017F}, if
the mass ratio is as small as 2:1 or 3:1, single events at a
distance of 100~Mpc could reveal the neutron star radius to
within 10--50\% \cite{2012PhRvD..85d4061L,2013arXiv1303.6298L},
depending on details.  This is a case in
which a combination of gravitational wave and electromagnetic
observations, along with simulations, would work very well: the
gravitational wave observation identifies the masses of the black
hole and neutron star and the spin of the black hole, and the
electromagnetic observation plus simulations derives information 
about the remaining mass of the disk (particularly if the recent promising observations results related to kilonovae hold up; see \cite{2013ApJ...774L..23B,2013Natur.500..547T}).  Simulations are still very
much in their infancy, but this is a promising avenue to pursue.
It has also been noted that the precision of constraints will be
improved significantly by combining the analyses of $>10$ bursts,
if systematic errors are under control \cite{2013PhRvL.111g1101D}, and
that observations with the planned third-generation gravitational
wave detector the Einstein Telescope will improve precision by
an order of magnitude \cite{2013arXiv1303.7393V}.

In summary, observations of gravitational waves from compact object
coalescences will yield promising new constraints on the properties
of neutron stars.  This will occur because of new mass measurements
and radius constraints.  Most of the information exists at high
frequencies, so in order to maximize this information it may be
necessary to explore techniques beyond the second generation of
detectors, such as squeezing of light \cite{2013NaPho...7..613A}.

\section{Summary}
\label{sec:summary}

Significant progress has been made in the last decade on mass estimates
of neutron stars.  The maximum
mass is $\gta 2~M_\odot$, which along with the lack of evidence of
rapid cooling suggests that non-nucleonic degrees of freedom appear
unnecessary to explain current data.  There is, however, considerable
freedom that would allow exotic phases.  The main limitation of 
existing observations is that radius estimates are 
shrouded in systematic
errors.  No current method is both precise and reliable
enough to pose significant constraints on the structure of neutron
stars.  This is unfortunate, because good radii would do more than
any other single measurement to inform us about the matter in the
cores of neutron stars \cite{2001ApJ...550..426L}.

Where, then, do we stand?  In the near future it is plausible that
our understanding of thermonuclear X-ray bursts, accreting neutron
stars, or the emission from cooling neutron stars may evolve to
the point that the complex phenomena associated with them can be
interpreted with confidence, and reliable radii and
masses will emerge.  In principle the needed
data have already been collected, but our evolving understanding that,
e.g., bursts do not settle quickly to a constant area of emission
and that cooling neutron stars display nonthermal emission has led
to caution.  At the same time, numerical simulations are improving
rapidly in sophistication.  It is possible, although far from
guaranteed, that within a few years magnetohydrodynamic simulations of accretion disks or bursting atmospheres will yield results
that are close enough to what is observed that our understanding
will solidify and radius measurements will become a powerful tool
for constraining the state of matter in the cores of neutron stars.

If ground-based gravitational wave facilities make detections as
expected within five years, then as we discussed in the previous
section this will open up a new way to measure masses and radii,
with systematic errors that are at a minimum different from those
that bedevil current attempts, and more optimistically might be
less significant than statistical uncertainties.
However, most of the information resides at high frequencies
where, at least for standard configurations, second-generation 
detectors will be insensitive enough that it may require a rare
high-signal event to derive restrictive constraints.  A third-generation
detector such as the Einstein Telescope should be able to obtain
all the required information, and even before such detectors exist it
may be possible to use configurations optimized for high frequencies
or 2.5 generation technology such as squeezed light to obtain the
information.

Electromagnetic observatories are also improving, and some of the
old reliable methods may improve our understanding substantially.
For example, the Jansky Very Large Array or (in roughly a decade)
the Square Kilometer Array might have enough sensitivity, bandwidth,
and computer power to detect a much larger population of double neutron
star binaries, among which we might by chance have some with stars
of $M>2.0~M_\odot$.  Even without such serendipitous discoveries,
the mere accumulation of time and data on NS-WD binaries in globular
clusters seems likely to yield high-mass objects that will provide
firm lower limits to the maximum mass that are much stronger than
currently exist.  As we have discussed, many other improvements
are expected using the large area and excellent spectral resolution of
ATHENA+ \cite{2013arXiv1306.2307N}, and the high area and timing
resolution of NICER \cite{2012SPIE.8443E..13G} and LOFT \cite{2012ExA....34..415F}.  This is especially true of radius estimates from fits to X-ray waveforms.

Overall, although we expect that eventually radius measurements will play a major role, in the next several years it appears that mass
measurements of neutron stars in binaries will continue to dominate
the discussion of the cold high-density equation of state.  The
expected improvements in data and models will allow
us to provide nuclear physicists with more certain constraints, and
we await the theoretical developments that result.

\begin{acknowledgement}
This work was supported in part by NSF grant AST0708424, by NASA ATP
grants NNX08AH29G and NNX12AG29G, and by grant number 230349 from the Simons Foundation.  We appreciate helpful suggestions from
Didier Barret, Paulo Bedaque, Sudip Bhattacharyya, David Blaschke, Tom
Cohen, Peter Jonker, Fred Lamb, Jim Lattimer, Ilya Mandel, Dany Page,
Bettina Posselt, Scott Ransom, Ingrid Stairs, and Natalie Webb.
\end{acknowledgement}

\bibliography{eosreview}

\end{document}